%% file: apssamp.tex
\documentclass[preprint,prd,amsfonts,showpacs, longbibliography,nofootinbib,amsmath,amssymb
]{revtex4-2}
\usepackage{bm,graphics,graphicx,epsfig,soul,latexsym,hyperref, float,mathrsfs,multirow,comment,amsmath,latexsym,appendix}
\usepackage[usenames]{xcolor}
\usepackage{tikz}
\usetikzlibrary{decorations.pathmorphing,decorations.markings}
\usepackage{dblfloatfix}
\usepackage{enumitem}

\graphicspath{{figure/}{./}}
\usepackage[english]{babel}
\usepackage[normalem]{ulem}
\definecolor{darkgreen}{rgb}{0,0.5,0}

\hypersetup{
    bookmarks=true,         
    unicode=false,          
    pdftoolbar=true,        
    pdfmenubar=true,        
    pdffitwindow=false,     
    pdfstartview={FitH},    
    pdftitle={My title},    
    pdfauthor={Author},     
    pdfsubject={Subject},   
    pdfcreator={Creator},   
    pdfproducer={Producer}, 
    pdfkeywords={keyword1} {key2} {key3}, 
    pdfnewwindow=true,      
    colorlinks=true,       
    linkcolor=red,          
    citecolor=cyan,        
    filecolor=magenta,      
    urlcolor=darkgreen,}

\begin{document}


\title{Worldline EFT treatment of quadratic and cubic gravity theories}

\author{Raghotham A Kulkarni}
 \email{raoulskulkarni@gmail.com}
 \affiliation{Center for High Energy Physics, Indian Institute of Science, Bangalore 560012, India}

\author{Rahul}
 \email{rahuldahiya98133@gmail.com}
\affiliation{Tata Institute of Fundamental Research, Mumbai, 400005, 
India}
\author{Soham Bhattacharyya} 
\email{xeonese@gmail.com}
\affiliation{Department of Physics, Indian Institute of Technology Madras, Chennai 600036, India}
\affiliation{Centre for Strings, Gravitation and Cosmology, Department of Physics, Indian Institute of Technology Madras, Chennai 600036, India}

\author{Dawood Kothawala}
\email{dawood@iitm.ac.in}
\affiliation{Department of Physics, Indian Institute of Technology Madras, Chennai 600036, India}
\affiliation{Centre for Strings, Gravitation and Cosmology, Department of Physics, Indian Institute of Technology Madras, Chennai 600036, India}

\begin{abstract}
This paper explores modifications to General Relativity (GR) by considering higher-order curvature terms in the gravitational action, specifically focusing on the quadratic Ricci scalar and a particular cubic contraction of the Riemann tensor. These modifications introduce new interactions at short distances, potentially altering the dynamics of compact objects. We calculate the effective two-body binding potential energy for these modified theories to quantify these effects using the worldline effective field theory (WEFT) formalism. This approach allows us to systematically integrate out short-distance gravitational effects, capturing the modifications to the binding potential. Our results demonstrate how the quadratic Ricci scalar and cubic Riemann tensor terms contribute to the two-body interaction at the leading order, highlighting deviations from classical GR predictions. These findings offer insight into the potential observational signatures of modified gravity theories in binary systems and other astrophysical settings.
\end{abstract}

\maketitle

\section{Introduction and Motivation}\label{sec:intro}\sloppy
Ever since \cite{Goldberger:2004jt} developed a novel method for describing the dynamics of General Relativity (GR) using the Worldline Effective Field Theory (WEFT), interest in it has spiked. Using the path integral approach of Quantum Field Theory (QFT), one can now study both conservative and radiative dynamics of the binary problem in gravity. While the calculations of the binary problem have progressed significantly in the standard post-Newtonian (PN) formalism, they remain slow and cumbersome. WEFT of GR has delivered results for conservative and radiative dynamics in a few years, compared to what it took for the PN formalism to achieve over several decades. This efficiency has attracted a significant amount of attention to the field of WEFTs in recent years. Many reviews of the classical aspects of WEFTs have come up, which include \cite{Goldberger:2007hy,Porto:2016pyg,Levi:2018nxp,Brunello:2022zui}, as well as quantum gravity effects like \cite{Donoghue:1994dn,Donoghue:1995cz,Donoghue:2017ovt,Donoghue:2012zc,Donoghue:2017pgk,Burgess:2003jk}.\\\\
Since the WEFT approach to GR has been tackled to a very high PN order, the next logical step is modifying the gravitational action. Modified theories of gravity have gained significant traction since the early days of GR, when it became evident that GR was not a complete theory of gravity. The singularity at the centre of a black hole (BH), the flattening of rotation curves of galaxies, the cosmological constant problem, and others have plagued GR. Two approaches to dealing with the problems of GR have emerged: string theory and loop quantum gravity. At the classical limit, both theories are expected to reproduce GR. This study will deal with the first relevant subdominant classical effect of a possible high-energy theory of gravity.\\\\
GR is obtained from the Einstein-Hilbert (EH) action defined as follows
\begin{equation}\label{eq:EH0}
    S_{EH} = \frac{c^4}{16\pi G} \int d^4x \sqrt{-g} R
\end{equation}
where $R$ is the Ricci Scalar. In the renormalization program of GR, it was found by \cite{tHooft:1974toh} that GR was renormalizable at one loop order. It was found that the divergences could be encoded in curvature squared terms like $R^2, R_{\mu\nu} R^{\mu\nu}, R_{\mu\alpha\rho\beta} R^{\mu\alpha\rho\beta}$. However, it was also found in \cite{tHooft:1974toh} and by later authors that the curvature squared term could be field redefined away by a suitable metric redefinition. Hence, it is not prudent to include curvature squared terms for a proper theory of modified gravity in the absence of matter or in BH spacetimes. However, as a proof of concept, we will show the leading order change in the two-particle effective potential for the Ricci squared term added to the EH action.\\
\\It has been known for quite some time that the S matrix of GR is divergent (non-renormalizable) at two-loop order in four dimensions \cite{Goroff:1985sz,Goroff:1985th}. The two loop divergences correspond to curvature cubed terms (scalars) and quadratic scalars of the covariant derivative of Ricci scalars and tensors. Utilising the field equations makes most of the terms vanish except for two Riemann cubed terms, of which only one is independent. Therefore, all two-loop divergences of the EH action can be parametrized by the following term
\begin{equation}\label{eq:rencubed}
    \Gamma_\infty^{(2)} = \sqrt{-g} \, R^{\alpha\beta}_{\quad\gamma\delta} R^{\gamma\delta}_{\quad \mu\nu} R^{\mu\nu}_{\quad \alpha\beta}.
\end{equation}
While the particular curvature cubed term of Eq. (\ref{eq:rencubed}) appears in the renormalization program as an actual divergence that cannot be renormalized, at a more recent time, a different procedure of field redefinition led to the same result. In \cite{deRham:2020ejn}, it was shown that field redefinitions of the metric tensor could get rid of all quadratic terms, as well as all but two Riemann curvature cubed terms, from an action containing higher-derivative terms added to the EH action as an EFT. It was further shown in the similar spirit of \cite{Goroff:1985sz,Goroff:1985th} by \cite{Cano:2019ore} that out of the two Riemann cubed terms, only one is independent and is again given by 
\begin{equation}\label{eq:frcubed}
    \sqrt{-g} \, \lambda_{ev} \,R^{\alpha\beta}_{\quad\gamma\delta} R^{\gamma\delta}_{\quad \mu\nu} R^{\mu\nu}_{\quad \alpha\beta}.
\end{equation}
where $\lambda_{ev}$ is some constant parametrizing the strength of the term in the action. This Riemann cubed term of Eq. (\ref{eq:frcubed}) cannot be field redefined away. Thus, two seemingly different approaches seem to lead to the same Riemann cubed term, implying a connection between the renormalization program and field redefinition strategies. Hence, when added to the EH action, the Riemann cubed term can be a candidate for the first subdominant effect of a low-energy limit of a high-energy theory of gravity in vacuum. Therefore, in this study, we find the two-particle effective potential, using the WEFT formalism, of the EH action plus the Riemann cubed term, that is
\begin{equation}
    S = \frac{c^4}{16\pi G} \int d^4x \sqrt{-g} \left(R + \alpha \, R^{\alpha\beta}_{\quad\gamma\delta} R^{\gamma\delta}_{\quad \mu\nu} R^{\mu\nu}_{\quad \alpha\beta}\right) \label{modact1}
\end{equation}
where $\alpha$ here is a fundamental constant that quantifies the cubic term's strength. Non-rotating black hole solutions for this theory were found in \cite{deRham:2020ejn}. The slowly rotating solution (along with the cubic odd parity part) was found in \cite{Cano:2021myl}, and perturbations of the rotating solution, \textcolor{black}{as well as corrections to the quasinormal modes (to the first order in spin) were studied} in \cite{Cano:2023tmv}. The Quasi-normal mode frequencies for the non-spinning case, as well as the odd parity part, have recently been found in \cite{Silva:2024ffz} for this particular theory. \textcolor{black}{Following this, Ref. \cite{Cano:2024ezp} found the corrections to the quasinormal mode overtones for the spinning geometry.} Recently, Ref. \cite{Glavan:2024cfs} studied the degrees of freedom in this model and found that the theory only carries a spin-2 field but features a preferred frame at short distances, suggesting a possible connection to unknown ultraviolet physics.\\

{\color{black} A post-Minkowskian study using modern scattering amplitude techniques was performed on a variant of the action (\ref{modact1}) in Refs. \cite{Brandhuber:2019qpg,Emond:2019crr} where it was argued that in non-vacuum space-times, there exists another independent combination of Riemann cubed terms which can contribute to the effective potential. However, as was mentioned in \cite{Brandhuber:2019qpg}, the three and four point graviton amplitudes of the extra combination vanishes. However, it was also argued that the extra combination can affect tidal terms at 5 PN order. However, in this study, we consider only BH space-times and therefore will not consider the additional combination. The above Refs. calculated the conservative dynamics of a compact binary system. Radiative dynamics of the variant of the action (\ref{modact1}) was calculated in Ref. \cite{AccettulliHuber:2020dal}. Finally, in Ref. \cite{Liu:2024atc}, constraints were placed on the coefficient $\alpha$ and the coefficient of the extra combination using the GW170608 event. \\

}

This paper is structured as follows: In Sec. \ref{sec:review}, we briefly review the WEFT formalism and reproduce the effective binding potential energy for a BBH system until the second post-Newtonian (PN) order in GR. In Sec. \ref{sec:quadgrav}, we derive the leading order change in the two-particle effective potential binding energy due to adding the Ricci squared term to the EH action. We obtain the same for the Riemann cubed gravity in Sec. \ref{sec:cubegrav}. Finally, in Sec. \ref{sec:concanddiss}, we conclude this study by identifying future directions and discuss the implications of the results of the current study to the two body problem of gravity.\\

We have worked with $\hbar = 1$ units throughout the text. We use the mostly plus signature of the Minkowski metric.

\section{Worldline EFT of GR: a mini review}\label{sec:review}
Worldline effective field theories (EFTs) of gravity are a powerful framework used to study the dynamics of massive objects, such as black holes or neutron stars, in GR and modified theories of gravity. They combine quantum field theory and general relativity techniques to simplify the complex gravitational interactions between massive bodies. This particular framework has proven to be highly efficient in obtaining various PN results for GR compared to the standard PN formalism, which required decades of research. The word `effective' in EFTs of gravity or other QFTs implies that an approximation is being made when it comes to the scales (which can be energy, for example) of the problem. In theories which have phenomena covering a wide range of energy or length scales and where the various scales are decoupled from each other, EFTs are used to calculate the effects of a different energy (length) scale, usually higher, on the energy (length) scale in question, usually lower. In the case of gravity, short (long) distance effects are synonymous with high (low) energies.\\

The compact binary problem of gravity is one such problem where there is a clear separation of scales at the early in-spiral, or when the factor $\frac{v}{c}$ ($v$ being the characteristic orbital velocity) is close to zero (non-relativistic). In such a situation, the hierarchy of scales becomes 
\begin{equation}\label{eq:scales}
    d << r << \lambda
\end{equation}
where $d$ is the length scale of the compact objects in question, $r$ is the relative separation between them, and $\lambda \sim \frac{r}{v}$ is the wavelength of GW radiation due to the time-varying quadrupolar nature of the binary system. It is to be noted that the above scales become comparable to each other at the late inspiral stage, and the whole formalism (similar to PN) breaks down.\\\\
In the hierarchy of scales, we first encounter physics in the region of one of the compact objects we are interested in. A compact object, at the first approximation, can be thought of as a point particle whose centre of mass is moving along a worldline $x^\mu(\lambda)$ that is coupled to the gravitational field $g_{\mu\nu}$, where $\lambda$ is an arbitrary affine parameter acting as some time, or can be taken as the proper time along its trajectory. However, real compact objects in nature are not point particles but have some `spread'. This spread can be perturbatively treated as multipole moments of the source energy-momentum tensor. This requires an in-depth knowledge of the source energy-momentum tensor. However, we generally do not have complete information on the energy-momentum tensor. In such a case, we can parametrize our ignorance about the spread of the compact object as a series of operators with Wilson coefficients proportional to the so-called `Love numbers' of the compact object. This spread and the Love numbers are directly related to the tidal response of one of the compact objects in the presence of the other. Thus, the effective point particle action can be written as follows
\begin{equation}\label{eq:Sppeff}
    S_{pp} = -m c^2 \int d\tau + c_S \int d\tau R + c_T \int d\tau R_{\mu\nu} \dot{x}^\mu \dot{x}^\nu + c_E \int d\tau E_{\mu\nu}E^{\mu\nu} + c_B \int d\tau B_{\mu\nu} B^{\mu\nu} + \cdots
\end{equation}
where in the first term $d\tau = \sqrt{g_{\mu\nu} dx^\mu dx^\nu}$, and is the result of the geodesic motion with respect to the metric $g_{\mu\nu}$. The second and third terms with Wilson coefficients $c_{S/T}$ can be removed by a field redefinition of the metric, as argued in \cite{Goldberger:2004jt}. Therefore, they can be ignored. The fourth and fifth terms contain contractions of the so-called electric and magnetic parts of the Riemann tensor, defined as follows:
\begin{equation}\label{eq:elec&magn}
    E_{\mu\nu} = R_{\mu\nu\alpha\beta} \dot{x}^{\alpha} \dot{x}^{\beta} \qquad B_{\mu\nu} = \epsilon_{\mu\alpha\beta\rho} R^{\alpha\beta}_{\sigma\nu} \dot{x}^{\sigma}
\end{equation}
However, since we are only interested in BBH space-times in this study, the constants $c_E$ and $c_B$ also vanish in GR, as was first shown in {\color{black} Refs. \cite{Damour:2009vw,Binnington:2009bb} and was reconfirmed in Ref. \cite{Ivanov:2022qqt} using scattering amplitude techniques}. \textcolor{black}{However, this is not in general true for modified gravity theories, especially for the cubic Riemann theory that we consider in this study, as were shown in Refs. \cite{DeLuca:2022tkm,Charalambous:2022rre}. But it should also be noted that tidal terms are PN suppressed, and that at the leading order contributions due to the cubic Riemann term will be only due to the point particle minimal coupling to gravity, which is what we consider in this study.} Hence, we are simply left with the first term, that is
\begin{equation}\label{eq:pp}
    S_{pp} = -m c^2 \int d\Lambda d^4x \delta^4\left[x^\mu-x^\mu_{pp}(\Lambda)\right] \sqrt{g_{\mu\nu} v^\mu v^\nu}
\end{equation}
where $x^\mu_{pp}(\Lambda)$ is the worldline of one of the point particles parametrized by $\Lambda$ and $v^\mu = \frac{dx^\mu}{d\Lambda}$. \\

A step up from the smallest length scale, $d$, is the near zone, $r$. In this zone, the weak field regime applies, and one can expand the metric tensor about the flat spacetime $\eta_{\mu\nu}$ as
\begin{equation}\label{eq:metpert}
    g_{\mu\nu} = \eta_{\mu\nu} + h_{\mu\nu}
\end{equation}
Therefore, to obtain the effective two-particle PN action (and hence the Lagrangian or the effective binding potential), one needs to integrate out the graviton $h_{\mu\nu}$ in the following manner
\begin{equation}\label{eq:pathintegral}
    e^{i S_{eff}} = \int Dh_{\mu\nu} e^{i S_{EH}[h] + i S_{GF}[h] + i S_{pp}[h,x^\mu]}.
\end{equation}
where
\begin{equation}\label{eq:EH1}
    S_{EH} = \frac{c^4}{16\pi G} \int d^4x \sqrt{-g} R[h,\partial h,\partial^2 h]
\end{equation}
is the Einstein-Hilbert action and 
\begin{equation}\label{eq:GF}
    S_{GF} = -\frac{c^4}{32\pi G} \int d^4x \sqrt{-g} \Gamma_\mu \Gamma^\mu
\end{equation}
where $\Gamma^\mu = \Gamma^\mu_{\alpha\beta} g^{\alpha\beta}$ is the contracted Christoffel symbol and $S_{GF}$ is known as a gauge-fixing term. We use the harmonic gauge throughout this study.\\

To fully exploit the separation of scales in the binary problem, one usually splits further the perturbation $h_{\mu\nu}$ into potential $(H_{\mu\nu})$, corresponding to the short length scale, and radiation $(\bar{h}_{\mu\nu})$ modes, corresponding to the larger length scale, as was done in {\color{black} Ref. \cite{Goldberger:2004jt} and was further explored in the reviews \cite{Porto:2016pyg,Brunello:2022zui}}. This separates the physics of $r$ from the physics of $\lambda$. The potential modes are treated as near instantaneous, given that the rate of change of the configuration of the binary is much slower than the rate at which gravitons are exchanged between the two BHs. To get the conservative dynamics, or the binding potential of the binary, one needs to integrate out both the potential and radiation mode, that is
\begin{equation}\label{eq:doubleintegral}
    e^{i S_{eff}[x^\mu]} = \int D \bar{h}_{\mu\nu} \int D H_{\mu\nu} e^{i S_{tot} (x^\mu, \bar{h}, H)},
\end{equation}
where $S_{tot}$ was defined in the RHS of Eq. (\ref{eq:pathintegral}).\\

Given the nonlinearity of the net action in the RHS of Eq. (\ref{eq:pathintegral}) and (\ref{eq:doubleintegral}), one must perform the integral perturbatively using Feynman diagrams. Once the effective action has been calculated, the real part of the action will lead to the conservative Lagrangian of the two-point particle system with the graviton integrated out. Similarly, the imaginary part of the effective action encodes the system's dissipative part or is proportional to the total radiated power through GW emission. However, in this paper, we will restrict ourselves to the conservative part of the action only and deal with the radiative effects in a follow-up paper.\\

{\color{black} This paper, however, uses a Kaluza-Klein decomposition of the potential modes putting the radiation modes to zero since we are dealing only with the conservative part of the problem}. Given a d+1 dimensional spacetime, we can rewrite the metric components, as was done in \cite{Kol:2007bc}, as follows
\begin{equation}\label{eq:KKdecomp}
    g_{\mu\nu}=e^{2\phi}\Biggl(\begin{matrix}
 -1& A_j \\
A_i & e^{-c_d\phi}\gamma_{ij}-A_iA_j
\end{matrix} \Biggr)\ , \qquad  \qquad \gamma_{ij}=\left(\delta_{ij}+\sigma_{ij}\right) \ .
\end{equation}
where $c_d = 2 \left(\frac{d-1}{d-2}\right)$ and $\delta_{ij}$ is the flat d dimensional metric. Using the full metric in Eq. (\ref{eq:KKdecomp}) and the metric perturbation equation (\ref{eq:metpert}), we can read off the metric perturbation $h_{\mu\nu}$ as
\begin{equation}\label{eq:KKhdecomp}
 h_{\mu\nu} =  \Biggl(\begin{matrix}
 1 - e^{2\phi} & A_j \\
A_i & e^{(2-c_d)\phi}\gamma_{ij} - e^{2\phi} A_iA_j - \delta_{ij}
\end{matrix} \Biggr)
\end{equation}
where $\phi$ is a scalar field, $A^i$ is a d-dimensional vector field, and $\sigma_{ij}$ is a $d\times d$ symmetric tensor field. Given the three (scalar, vector, and tensor) fields, Eq. (\ref{eq:pathintegral}) or (\ref{eq:doubleintegral}) gets modified into the following
\begin{equation}\label{eq:KKpathintegral}
    e^{i S_{eff}(x^\mu)} = \int D\phi DA_i D\sigma_{jk} e^{iS_{EH}[\phi,A_i,\sigma_{jk}]+iS_{pp}[x^\mu,\phi,A_i,\sigma_{jk}]+iS_{GF}[\phi,A_i,\sigma_{jk}]}
\end{equation}
We obtain the interaction terms for the perturbative path integral treatment by expanding the three terms of the action $(S_{EH}, S_{pp}, S_{GF})$ till the {\color{black} sixth } order in $h_{\mu\nu}$ and then decompose $h$ into $\left(\phi, A_i,\sigma_{jk}\right)$, as shown in Eq. (\ref{eq:KKhdecomp}). We also scale the fields $\left(\phi, A_i,\sigma_{jk}\right)$ in powers of $c$ to obtain a PN series of the interaction terms in the bulk and worldline action in the following manner
\begin{eqnarray}\label{eq:scaling}
    \phi &\rightarrow& \frac{\phi}{c^2} \\
    A_i &\rightarrow& \frac{A_i}{c^3} \\
    \sigma_{jk} &\rightarrow& \frac{\sigma_{jk}}{c^4}.
\end{eqnarray}
{\color{black} The rationale behind this particular type of scaling is a lesson that is learnt from the PN expansion of GR. It is well known (for example, see Ref. \cite{Maggiore:2007ulw}) that the time-time component of the metric perturbation in PN theory scale as $\frac{1}{c^2}$, the time-space component as $\frac{1}{c^3}$, and the space-space component as $\frac{1}{c^4}$. Hence, we scale the corresponding Kaluza-Klein fields in the same manner. }\\

After this, we perturbatively evaluate Eq. (\ref{eq:KKpathintegral}) by first identifying the appropriate two-point correlation functions (or the propagators) and the interaction vertices of the fields $\left(\phi, A_i,\sigma_{jk}\right)$ from the expansion of the full action in Kaluza-Klein (Kol-Smolkin) variables and by transforming to the Fourier domain. The path-integral is computed diagrammatically, order by order, characterized by a definite scaling of the speed of light $c$ and Newton's constant $G$. Since we first look for the non-relativistic limit (or slow motion of the sources), we can take $\frac{k_0}{|\Vec{k}|}$ as small and expand the propagators around $\frac{k_0}{|\Vec{k}|} = 0$. For the exact form of the full propagators, see Appendix A of Ref. \cite{Brunello:2022zui}.\\

We begin by drawing all possible Feynman diagrams, carefully accounting for the scaling and order of the vertices. The total amplitude is the sum of all the topologically distinct and connected diagrams. After evaluating the amplitude and by performing the momentum integrals for the internal lines, the inverse Fourier transform of the amplitude (after putting $d = 3$) leads to a time integral $\int\mathcal{L}_{eff}\, dt$ from which we extract the effective Lagrangian $\mathcal{L}_{eff}$, and eventually the effective potential $V_{eff}$. {\color{black} The full effective Lagrangian at various PN orders till 2 PN, including the kinetic terms, have been given in Appendix \ref{app:CoM}. Since we do not change the worldline action due a change in the theory of gravity, the kinetic terms remain the same as in GR. } \\

{\color{black} We find that after including the sixth order of metric perturbation, the Kaluza-Klein expanded action stops changing, implying that inclusion of the sixth order of metric perturbation is enough to obtain the correct metric expanded action till the second PN order at least. Similarly, we have found that the 2 PN two-particle effective Lagrangian also does not change after including the sixth order of perturbation of the metric in the action, and reproduces the same as in previous literature in Refs. \cite{deAndrade:2000gf,Bernard:2017ktp}, where they quote effective Lagrangian. We reproduce the two-particle effective binding potential energy $V_{eff}$ till 2 PN in this paper for the benefit of the reader as follows:
\begin{eqnarray}\label{eq:effpot}
    V_{eff} &=& V_{N} + \frac{1}{c^2} V_{1PN} + \frac{1}{c^4} V_{2PN} \\
    V_N &=& -\frac{G m_1 m_2}{r} \\
    V_{1PN} &=& -G m_1 m_2 \left[\frac{\left\{3 v_1^i v_{1,i}+\left(3 v_2^i-7 v_1^i\right) v_{2,i}\right\}}{2 r}-\frac{ v_1^i x_i v_2^{j} x_{j}}{2 r^3}\right] + \frac{G^2
   m_1 m_2 \left(m_1+m_2\right)}{2 r^2} \nonumber \\
   \\
   V_{2PN} &=& -G m_1 m_2 \left[-\frac{1}{8 r^3}\left\{ v_1^i \left[v_1^{j} x_{j} \left\{x_{k} \left(6 v_2^{k} v_{1,i}-12 v_2^{k} v_{2,i}-a_2^{k} x_i\right)+x_i v_2^{k}
   v_{2,k}\right\}+v_2^{j} v_2^{k} x_{k} \left(6 x_i v_{2,j}+x_{j} v_{1,i}\right)\right]\right.\right.\nonumber\\
   &&\left.+a_1^i x_i v_2^{j} x_{j} v_2^{k} x_{k}\right\}+\frac{1}{8 r}\left\{a_1^i x_i v_2^{j} v_{2,j}-14 a_1^i v_2^{j} x_{j} v_{2,i}+v_1^i v_{1,i} \left(7 v_1^{j} v_{1,j}-10 v_1^{j} v_{2,j}+3 v_2^{j}
   v_{2,j}\right.\right.\nonumber\\
   &&\left.-a_2^{j} x_{j}\right)+2 v_1^i \left[v_2^{j} \left(-6 x_i a_{2,j}+6 x_{j} a_{1,i}-5 v_{2,i} v_{2,j}\right)+v_1^{j} \left(7 x_{j} a_{2,i}+v_{2,i}
   v_{2,j}\right)\right]+7 v_2^i v_2^{j} v_{2,i} v_{2,j}\nonumber\\
   &&\left.\left.-a_1^i a_2^{j} x_i x_{j}\right\}+\frac{15}{8} r a_1^i a_{2,i}+\frac{3  v_1^i x_i v_1^{j} x_{j}
   v_2^{k} x_{k} v_2^{l} x_{l}}{8 r^5}\right] - G^2 m_1 m_2 \left[\frac{1}{4 r^2}\left\{m_1 \left[8 v_1^i v_{1,i}+7 \left(v_2^i-2 v_1^i\right) v_{2,i}\right]\right.\right.\nonumber\\
   &&\left.\left.+m_2 \left(7 v_1^i v_{1,i}+2 \left(4 v_2^i-7 v_1^i\right) v_{2,i}\right)\right\}+\frac{1}{2 r^4}\left\{x_i x_{j} \left(m_1 v_2^i
   v_2^{j}+m_2 v_1^i v_1^{j}\right)\right\}\right] \nonumber\\
   && - \frac{G^3 m_1 m_2 }{2 r^3}\left(m_1^2+6 m_2 m_1+m_2^2\right)
\end{eqnarray}
where $r = \sqrt{x^i x_i}$ is the distance between the two point particles, $x^i$ is the separation vector, $a^i_K$ for $K = 1,2$ is the acceleration vector and $v^i_K$ for $K = 1,2$ is the velocity vector for the first and the second object. We only provide the effective potential and not the PN kinetic energy, since it remains unchanged under changes in the bulk action.

\section{Proof of concept: quadratic gravity}\label{sec:quadgrav}
We start by adding the Ricci scalar squared to the EH action and by fixing the gauge to harmonic. The full action in the harmonic gauge is then given in d = 3 as follows
\begin{eqnarray}\label{SQuad}
    S_{quad} &=& S_{bulk} + S_{pp} \\
    S_{bulk} &=& \frac{c^4}{16 \pi G } \int d^4x \,\sqrt{-g} \, (R  + \alpha R^2 - \frac{1}{2}\Gamma^\mu \Gamma_\mu)\\
    S_{pp} &=& - \sum_{a=1}^2 m_a c^2 \int d\tau_a  \label{eq:pp}
\end{eqnarray}
where $\Gamma^\mu$ was defined right after Eq. (\ref{eq:GF}), $m_a$ for $a = 1,2$ are the rest masses of two point particles moving on a wordline $x^\mu_a$. \\

While $S_{pp} $ remains the same as in GR, $S_{bulk}$ changes due to the inclusion of the $R^2$ term. The changes in the bulk action on perturbing the same and decomposing into Kaluza-Klein variables, as given in Eq. (\ref{eq:KKhdecomp}), are given till 1 PN as follows (we include the GR terms as well): 
\begin{eqnarray}\label{eq:RsquaredBulk}
S_{bulk} &=& \frac{1}{G}\int d^4 x \hspace{.1cm} \biggl[ -\frac{\left(\partial \phi\right)^2}{8 \pi} - \frac{\alpha\left(\nabla^2 \phi\right)^2}{4 \pi} + \frac{1}{c^2} \biggl\{ \frac{\partial_b A_a \partial^b A^a}{32 \pi} + \frac{\left(\partial_t \phi\right)^2}{8 \pi} + \alpha \biggl( -\frac{\phi \left( \nabla^2 \phi \right)^2 }{2 \pi} \nonumber\\
&&+ \nabla^2 \phi \biggl[ \frac{\partial_t \partial_a A^a}{2 \pi} + \frac{\left(\partial \phi\right)^2}{2 \pi} - \frac{\partial_b \partial_a \sigma^{a b}}{4 \pi} + \frac{\nabla^2 \sigma}{4 \pi} + \frac{3 \partial_t^2 \phi}{2 \pi} \biggr] \biggr) \biggr\} \biggr]
\end{eqnarray}
where $\sigma = \sigma^i_i$, $\nabla^2 \equiv \delta^{ij} \partial_i \partial_j$, $\delta_{ij}$ is the 3 dimensional Minkowski spatial metric, and $(\partial\phi)^2 = \partial_i \phi \partial^i \phi$. We also give the Kaluza-Klein decomposed point particle worldline action for one of the particles, till 2 PN, for the reader's convenience
\begin{eqnarray}
    S_{pp} &=& \int dt \left\{- m c^2 + \frac{1}{2} m v^2 - m \phi + \frac{m}{c^2}\left(\Vec{A}\cdot\Vec{v} + \frac{1}{8} |\Vec{v}|^4 - \frac{3}{2} |\Vec{v}|^2 \phi - \frac{1}{2} \phi^2\right) \right.\nonumber\\
    && \left.+ \frac{m}{c^4} \left[ \frac{1}{2} |\Vec{v}|^2 \Vec{A}\cdot\Vec{v} + \frac{1}{16} |\Vec{v}|^6 + \frac{1}{2} v^i v^j \sigma_{ij} + \left( \Vec{A} \cdot \Vec{v} - \frac{7}{8} |\Vec{v}|^4 \right)\phi + \frac{9}{4} |\Vec{v}|^2 \phi^2 -\frac{1}{6} \phi^3 \right] + \mathcal{O}\left(\frac{1}{c^5}\right) \right\}\nonumber\\
\end{eqnarray}

\subsection{Newtonian order or 0 PN}
At the Newtonian order, because of the particular scaling of Eq. (\ref{eq:scaling}), we only get $\phi$ and derivatives in Eq. (\ref{eq:RsquaredBulk}). Taking the Fourier transform of $\phi$ one can write $\phi(x)$ in terms of $\phi(k)$ as
\begin{eqnarray} \label{eq:fourierphi}
    \phi(x^\mu) &=& \int \frac{d^4k}{(2\pi)^4} e^{ik_\mu x^\mu} \phi(k^\mu) = \int \frac{d^4k}{(2\pi)^4} e^{i (\Vec{k} \cdot \Vec{x}  - c k_0 t)} \phi(k^\mu)
    \end{eqnarray}
In the non-relativistic limit, at the leading order, $k_0$ is ignored. The 0 PN diagram is just the exchange of one graviton (or the scalar part of the Kaluza-Klein metric decomposition) between the two worldlines and is given as follows.
\\

\hspace{6.5cm}\begin{tikzpicture}
    \draw (0,0) -- (2,0);
    \draw (0,2) -- (2,2);
    \draw (0,-0.05) -- (2,-0.05);
    \draw (0,2.05) -- (2,2.05);
    \draw (1,2) -- (1,0);
\end{tikzpicture}\\

{\color{black} In what follows, we will use double lines denote worldline factors, thin lines to denote the $\phi$ propagator, dashed lines to denote the $A^i$ propagator, and finally, wavy lines to denote the $\sigma_{ij}$ propagator. Also, a cross in the middle of a thin line implies that a factor of $\frac{k_0}{|\Vec{k}|}$ has been inserted in the $\phi$ propagator.} The above diagram corresponds to the leading order term in the series expanded generating functional $Z[J]$. The generating functional in quantum field theory encapsulates all correlation functions of a field theory and is defined through a path integral formulation, enabling the systematic computation of n-point functions via functional derivatives with respect to a source term \cite{stampPHYS508}. $Z[J]$ is defined as follows (since the diagram only involves $\phi$ we have suppressed the $A$ and $\sigma$ fields in the following)
\begin{equation}
    Z[J] = \int D\phi  e^{ i S[\phi(x^\mu)] + J(x^\mu) \phi(x^\mu) }
\end{equation}
where at the leading order,
\begin{equation}
    S[\phi(x^\mu)] \approx \frac{1}{8 \pi G} \int d^4x \left( \phi \nabla^2 \phi - \alpha \nabla^2 \phi \nabla^2 \phi \right)
\end{equation}
and $J(x^\mu)$ is the leading order source/interaction term coming from $S_{pp}$ and is given as follows
\begin{equation}
    J(x^\mu) =-i \sum_{a = 1}^{2}\int dt_a m_a c^2 \delta^4(x^\mu-x^\mu_a),
\end{equation}
where $x^\mu_a$ is the worldline of the $a$-th particle. As was performed in Ref. \cite{Zee:2003mt}, the `free' part of the action is kept in the exponential, while the source/interaction term (or the $J$ term) is expanded in a series. In the Fourier domain, the series expanded generating functional is given as follows
\begin{eqnarray}\label{eq:ZpartRsquaredNewt}
    Z[\tilde{J}] &\approx& \int D \phi(k) \,\,  e^{{-\frac{i}{8 \pi G} \int d^4k \phi(-k) (|\Vec{k}^2| +2\alpha |\Vec{k}|^4) \phi(k)}} \left(\frac{1}{2}\int d^4k_1d^4k_2 \tilde{J}(-k_1) \tilde{J}(-k_2)\phi(k_1)\phi(k_2) \right)\nonumber\\
\end{eqnarray}
where
\begin{eqnarray}
 \tilde{J}(k) = -i \sum_{a = 1}^{2}\int dt_a m_a c^2 e^{i k_\mu x^\mu_a}.
\end{eqnarray}
It is to be noted that although there is a double sum in Eq. (\ref{eq:ZpartRsquaredNewt}) owing to the product of the two $J$'s, we only take cross terms into account while performing the calculations and hence, we ignore the self energy terms.\\

The propagator is obtained at the leading order, and it is to be noted that it changes due to the term proportional to $\alpha$, and hence it changes the 2-point function as well
\begin{eqnarray}
    <\phi(k_1) \phi(k_2)> = \frac{- 4 \pi i G (2 \pi)^4\delta^4(k_1+k_2)}{(k_1^2 +2 \alpha k_1^4)}
\end{eqnarray}
We then calculate the amplitude for the one graviton exchange and take its inverse Fourier transform to obtain the effective potential, that is
\begin{eqnarray}\label{eq:YukawaPotential}
    V_{eff} &\propto& -\int d^4k_1d^4k_2 \tilde{J}(-k_1)\tilde{J}(-k_2)<\phi(k_1)\phi(k_2)> \\
            &=& -\frac{G m_1 m_2 }{r}\left(1- e^{\frac{-r}{\sqrt{2 \alpha}}}\right)
\end{eqnarray}
An important integration that we use in the calculation is 
\begin{eqnarray}
\int \frac{d^3k}{(2\pi)^3} \frac{e^{i \Vec{k}\cdot \Vec{r}_a}}{|\Vec{k}|^2+2 \alpha|\Vec{k}|^4} = \frac{1- e^{\frac{-r}{\sqrt{2 \alpha}}}}{4 \pi r}    
\end{eqnarray}
The result of Eq. (\ref{eq:YukawaPotential}) exactly matches what was found in Section 8 of \cite{Donoghue:2017pgk}. We can also consider $\alpha$ as a negative parameter; in this case, the above diagram gives
\begin{eqnarray}
    V_{eff} = -\frac{G m_1 m_2 }{r}\left[1- \cos{\left(\frac{r}{\sqrt{2 \alpha}}\right)}\right]    
\end{eqnarray}

However, if the length scale $\propto \sqrt{\alpha}$ is much smaller than the length scales in the potential zone, then the new pole of $R + \alpha R^2$ is outside of the cutoff of our approximation and hence the Yukawa like potential goes away. In the following, we show how field redefinition techniques can genuinely get rid of the quadratic corrections to the effective potential.

{\color{black}

\subsection{$\alpha$ as a perturbing parameter}

\subsubsection{Till 1 PN}

Eq. (\ref{eq:RsquaredBulk}), when treated as a perturbative bulk action in both $\frac{1}{c}$ and $\alpha$, can be simplified further by performing a field redefinition on $\phi$. Considering the following redefinition of $\phi$

\begin{equation}\label{eq:phiredefbulk}
    \phi \rightarrow \phi + \frac{2 \alpha}{c^2} \left( \partial \phi \right)^2 \, ,
\end{equation}

The field redefinition (\ref{eq:phiredefbulk}) changes the bulk action into the following

\begin{eqnarray}
    S_{bulk} &=& \frac{1}{G}\int d^4 x \hspace{.1cm} \biggl[ -\frac{\left(\partial \phi\right)^2}{8 \pi} - \frac{\alpha\left(\nabla^2 \phi\right)^2}{4 \pi} + \frac{1}{c^2} \biggl\{ \frac{\partial_b A_a \partial^b A^a}{32 \pi} + \frac{\left(\partial_t \phi\right)^2}{8 \pi} + \alpha \biggl( -\frac{\phi \left( \nabla^2 \phi \right)^2 }{2 \pi} \nonumber\\
&&+ \nabla^2 \phi \biggl[ \frac{\partial_t \partial_a A^a}{2 \pi}  - \frac{\partial_b \partial_a \sigma^{a b}}{4 \pi} + \frac{\nabla^2 \sigma}{4 \pi} + \frac{3 \partial_t^2 \phi}{2 \pi} \biggr] \biggr) \biggr\} \biggr]
\end{eqnarray}

which indicates that the $\nabla^2 \phi \left(\partial\phi\right)^2$ term gets canceled away by the field redefinition. Other $\alpha$ proportional terms lead to contact terms in the effective potential and do not lead to long-range interactions. However, now that the bulk has been modified, the worldline must also change. The worldline till 1 PN gets redefined as follows

\begin{eqnarray}
    S_{pp} &=& \int dt \left[- m c^2 + \frac{1}{2} m v^2 - m \phi + \frac{m}{c^2}\left\{\Vec{A}\cdot\Vec{v} + \frac{1}{8} |\Vec{v}|^4 - \frac{3}{2} |\Vec{v}|^2 \phi - \frac{1}{2} \phi^2 - 2 \alpha \left(\partial\phi\right)^2 \right\}   \right] \nonumber\\
\end{eqnarray}

However, it is to be noted that each of the fields appearing in the point particle action is evaluated at the worldline. Therefore, the point particle action can be written as

\begin{eqnarray}
    S_{pp} &=& \int dt \int d^4x \delta^4\left(x^\mu - x^\mu_{pp} (t)\right) \left[- m c^2 + \frac{1}{2} m v^2 - m \phi + \frac{m}{c^2}\left\{\Vec{A}\cdot\Vec{v} + \frac{1}{8} |\Vec{v}|^4 - \frac{3}{2} |\Vec{v}|^2 \phi - \frac{1}{2} \phi^2 \right.\right.\nonumber\\
    && \left.\left.- 2 \alpha \left(\partial\phi\right)^2 \right\}   \right]
\end{eqnarray}

where $x^\mu_{pp}(t)$ is the worldline of one of the point particles. Performing a second field redefinition

\begin{equation}
    \phi \rightarrow \phi \left[ 1 - 8 \alpha G \pi m \delta^4\left( x^\mu - x^\mu_{pp}(t) \right)  \right]
\end{equation}

leads to the cancellation of the $\alpha$ dependent term in the worldline against the opposite sign term in the bulk. This has been shown in Appendix A of Ref. \cite{Porto:2016pyg}. Extra terms coming in the worldline due to the second field redefinition changes the Green's function only by a overall normalization factor, and the physics remains the same, again as shown in Appendix A of Ref. \cite{Porto:2016pyg}. Hence, when $\alpha$ is treated as a small parameter, there is no correction to the Einstein-Infeld-Hoffmann (EIH) Lagrangian due to a quadratic correction to the EH action. In the following, we generalize the above method to all PN orders.

\subsubsection{For all PN orders}

One has the full unperturbed action of the quadratic Ricci scalar correction to the EH action coupled with to a point particle action as
\begin{eqnarray}
    S_{quad} &=& \frac{c^4}{16 \pi G} \int d^4x \sqrt{-g} \left( R + \alpha R^2 \right) - m c^2 \int d\tau \, d^4x \, \delta^4\left( x^\mu - z^\mu (\tau) \right) \sqrt{g_{\mu\nu} u^\mu u^\nu} 
\end{eqnarray}
Perturbing/varying the action with respect to the metric $g_{\mu\nu}$, that is
\begin{equation}
    g_{\mu\nu} = \bar{g}_{\mu\nu} + \epsilon h_{\mu\nu} \, ,
\end{equation}
for a background $\bar{g}_{\mu\nu}$ and a bookkeeping parameter $\epsilon$, and setting the variation $h_{\mu\nu}$ as 
\begin{equation}
    h_{\mu\nu} = -\frac{\alpha}{\epsilon} \,\, \bar{g}_{\mu\nu} \bar{R}
\end{equation}
leads to the cancellation of the Ricci scalar squared term in the bulk till $\mathcal{O}\left(\alpha\right)$, and leads to the following change in the net action
\begin{eqnarray}
    \bar{S}_{quad} &=& \frac{c^4}{16 \pi G} \int d^4x \sqrt{-\bar{g}} \bar{R} -m c^2 \int d\tau \, d^4x \, \delta^4\left( x^\mu - z^\mu (\tau) \right) \sqrt{\bar{g}_{\mu\nu} u^\mu u^\nu} \nonumber\\
    &&+ \frac{1}{2} m c^2 \alpha \int d\tau \, d^4x \, \delta^4\left( x^\mu - z^\mu (\tau) \right) \, \bar{R} \, \sqrt{\bar{g}_{\mu\nu} u^\mu u^\nu}
\end{eqnarray}
However, this is exactly as one would obtain instead by endowing the point particle action with a non-minimal coupling to the gravitational degrees of freedom in GR. As shown in Appendix A of \cite{Porto:2016pyg} and in \cite{Goldberger:2004jt}, a local field redefinition can eliminate such terms. Hence, when $\alpha$ is treated as a small parameter, there is no change to the effective action at any PN order.
}

\section{The two-particle effective Lagrangian of cubic gravity}\label{sec:cubegrav}
In this section, we try to calculate the leading order change in the two-body effective potential by integrating the gravitational degrees of freedom from the following action
\begin{eqnarray}
    S_{cubic} &=& S_{bulk} + S_{pp} \\
    S_{bulk} &=& \frac{c^4}{16 \pi G} \int d^4x \left(R + \alpha \, R^{\alpha\beta}_{\quad\gamma\delta} R^{\gamma\delta}_{\quad \mu\nu} R^{\mu\nu}_{\quad \alpha\beta} - \frac{1}{2} \Gamma^\mu \Gamma_\mu\right)
\end{eqnarray}
where $S_{pp}$ remains the same as in Eq. (\ref{eq:pp}). Like in the previous section, we use the harmonic gauge fixing term. Upon perturbing the metric and decomposing the metric perturbation as in Eq. (\ref{eq:KKhdecomp}), we find the following interaction terms in two different PN orders:

\subsection{PN-1}\label{subsec:1PNR^3}
At the first PN order, the bulk interaction terms for the cubic Riemann part are given as follows:
  \begin{align}
   \text{BIT} &= \left[\frac{3 \left(\partial_{i}\partial_{i}\phi\right) \left(\partial_{j}\partial_{l}\phi\partial_{j}\partial_{l}\phi \right)}{2G\pi}-\frac{ \left( \partial_{i} \partial_{j} \phi \, \partial_{j} \partial_{l} \phi \, \partial_{l} \partial_{i} \phi \right)}{ G \pi} \right],
  \end{align}
which correspond to the following diagrams\\

\hspace{3cm}\begin{tikzpicture}
\draw[postaction={decorate, decoration={markings, mark=at position 1 with {\fill[black] (0,0) circle (2pt);}}}] (2,1.7) -- (1,1);
   \draw (0,0) -- (0,2);
    \draw (2,0) -- (2,2);
    \draw (2,0.3) -- (1,1);
    \draw (0,1) -- (1,1);
    \draw (-0.05,0) -- (-0.05,2);
    \draw (2.05,0) -- (2.05,2);
\end{tikzpicture}\hspace{4cm}
\begin{tikzpicture}{h}
\draw[postaction={decorate, decoration={markings, mark=at position 1 with {\fill[black] (0,0) circle (2pt);}}}] (2,1) -- (1,1);
     \draw (0,0) -- (0,2);
    \draw (2,0) -- (2,2);
    \draw (0,1.7) -- (1,1);
    \draw (0,0.3) -- (1,1);
    \draw (-0.05,0) -- (-0.05,2);
    \draw (2.05,0) -- (2.05,2);
\end{tikzpicture}\\
{\color{black}
The Fourier integral corresponding to the first diagram is given by
\begin{align}
        I \propto \int _{k_1k_2k_3} \frac{k_1^2(k_2.k_3)^2}{k_1^2~k_2^2~k_3^2} \delta^4(k_3 - k_1 - k_2) e^{i(k_1+k_2).x_1-ik_3.x_2}
\end{align}
doing integral over $k_3$ and defining $p=k_1+k_2$
\begin{align}
   I \propto \int_{k_2} \frac{k_2^i~k_2^j}{k_2^2} ~ \int_p \frac{p^i~p^j}{p^2}e^{ip.r_{12}}=0
\end{align}

For the second diagram one obtains a vertex factor as $(k_1.k_2)(k_2.k_3)(k_3.k_1)$, which corresponds to the following integral
\begin{align}
    I &\propto\int _{k_1 k_2 k_3} \frac{(k_1.k_2)(k_2.k_3)(k_3.k_1)}{k_1^2~k_2^2~k_3^2} \delta^4(k_3-k_1-k_2) e^{i(k_1+k_2).x_1-ik_3.x_2}
\end{align}
Integrating over $k_1$ one obtains
\begin{align}
    I &\propto \int _{k_2 k_3} \frac{[(k_3-k_2).k_2](k_2.k_3)[k_3.(k_3-k_2)]}{(k_3-k_2)^2~k_2^2~k_3^2}e^{ik_3.r_{12}}
\end{align}
The above corresponds to four terms
\begin{align}
I &\propto\Bigg[\int _{k_2k_3} \frac{(k_2.k_3)^2}{(k_3-k_2)^2~k_2^2}e^{ik_3.r_{12}}-\int _{k_2k_3} \frac{k_2.k_3}{(k_3-k_2)^2}e^{ik_3.r_{12}}\nonumber\\
&-\int _{k_2k_3} \frac{(k_2.k_3)^3}{(k_3-k_2)^2~k_2^2~k_3^2}e^{ik_3.r_{12}} + \int_{k_2k_3} \frac{(k_2.k_3)^2}{(k_3-k_2)^2~k_3^2}e^{ik_3.r_{12}}\Bigg]
\end{align}
Using the results from the Appendix A of Ref. \cite{Levi:2011eq} we find that the integral is \(\propto \delta'''(r_{12}) \), where a prime corresponds to a time derivative, and which is zero for non-zero separation of the binary system. Hence, both diagrams evaluate to zero and one has no contribution at the first PN order. We move on to the second PN order.
}
\subsection{PN-2}
In the second PN order, we look at contributions from different Klauza-Klein fields appearing in the bulk interaction. We give the following: the type and kind of fields (or their mixtures) appearing in the interaction, the bulk interaction terms (BIT) themselves, the respective Feynman diagrams when coupled to the worldlines, {\color{black} and some of the calculation details regarding the integrals corresponding to the Feynman diagrams. We have extensively used the results of Appendix A of \cite{Levi:2011eq} in calculating the Fourier integrals corresponding to each diagram.}\\

$$\underline{\phi^4}$$
\begin{align}
\text{ BIT } &=\frac{-3 (\partial_i\partial^i\phi)^2~(\partial_j\phi \partial^j\phi)}{2G \pi}+\frac{3(\partial_i\partial^i\phi)~(\partial^j\phi \partial_k~\partial_j\phi\partial^k\phi)}{G\pi}+\frac{3 (\partial_i\phi \partial^i\phi)~(\partial_j\partial_k\phi~\partial^j\partial^k\phi])}{2G\pi}\notag\\
&-\frac{15(\partial_i\partial^j\phi~\partial^i\phi\partial^k\phi~\partial_j\partial_k\phi)}{2G\pi}
+\phi\left(\frac{6 \partial_k\partial^k\phi \partial_i\partial_j\phi\partial^i\partial^j\phi- 4 \partial_i\partial_j\phi\partial^j\partial_k\phi\partial^i\partial^k\phi}{G \pi}\right)
\end{align}
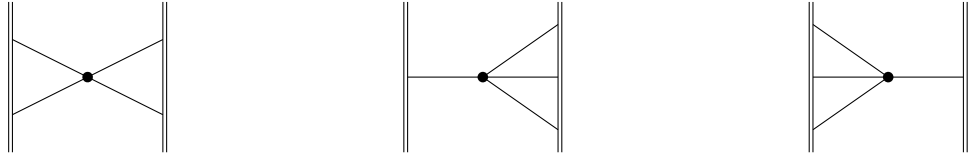
\hspace{1.5cm}
\begin{tikzpicture}{h}
\draw[postaction={decorate, decoration={markings, mark=at position 1 with {\fill[black] (0,0) circle (2pt);}}}](0,1.5) -- (1,1) ;
    \draw (0,0) -- (0,2);
    \draw (2,0) -- (2,2);
    \draw (0,0.5) -- (1,1);
    \draw (2,1.5) -- (1,1);
    \draw (2,0.5) -- (1,1);
    \draw (-0.05,0) -- (-0.05,2);
    \draw (2.05,0) -- (2.05,2);
\end{tikzpicture}\hspace{3cm}
\begin{tikzpicture}{h}
\draw[postaction={decorate, decoration={markings, mark=at position 1 with {\fill[black] (0,0) circle (2pt);}}}](0,1) -- (1,1) ;
    \draw (0,0) -- (0,2);
    \draw (2,0) -- (2,2);
    \draw (2,1.7) -- (1,1);
    \draw (2,0.3) -- (1,1);
    \draw (2,1) -- (1,1);
    \draw (-0.05,0) -- (-0.05,2);
    \draw (2.05,0) -- (2.05,2);
\end{tikzpicture}
\hspace{3cm}
\begin{tikzpicture}{h}
\draw[postaction={decorate, decoration={markings, mark=at position 1 with {\fill[black] (0,0) circle (2pt);}}}](2,1) -- (1,1) ;
    \draw (0,0) -- (0,2);
    \draw (2,0) -- (2,2);
    \draw (0,1.7) -- (1,1);
    \draw (0,0.3) -- (1,1);
    \draw (0,1) -- (1,1);
    \draw (-0.05,0) -- (-0.05,2);
    \draw (2.05,0) -- (2.05,2);
\end{tikzpicture}\\

{\color{black} A typical vertex factor for the \(\phi^4\) interaction arises from the four \(\phi\) interaction term in the Lagrangian. All momenta involved in the interaction are 3-vectors except for the delta function, which ensures momentum conservation. By integrating over \( k_1 \) using the delta function, we eliminate one of the momenta. Subsequently, performing the remaining integrals yields a result that is proportional to \( 1/r_{12}^7 \).

The expression for a typical vertex factor, along with the propagator and momentum conservation, is given by:
\begin{equation}
    \mathcal{V} = \frac{k_1^2 (k_2\cdot k_3)(k_3\cdot k_4)}{k_1^2 k_2^2 k_3^2 k_4^2}\delta^4( k_1+k_2+k_3+k_4)
\end{equation}
Hence, a typical integrand looks like
\begin{equation}
   \mathcal{I} = \frac{ k_1^2 (k_2\cdot k_3)(k_3\cdot k_4)}{k_1^2 k_2^2 k_3^2 k_4^2}\delta^4(k_1+k_2+k_3+k_4)\times e^{i(k_1 \cdot x_1)} e^{[i (k_2 +k_3 +k_4)\cdot x_2]}
\end{equation}
where \( k_1, k_2, k_3, k_4 \) are the 3-momenta of the vertex factor, the delta function \( \delta^4(k_1+k_2+k_3+k_4) \) enforces four-momentum conservation at the vertex, and the denominator accounts for the propagators of the intermediate particles.

After summing over all possible scalar product combinations corresponding to the $\phi^4$ diagrams, the final result takes the form:

\begin{equation}
   I =  \frac{160 G^3 \alpha( m_1 m_2^3 +m_1^3 m_2) }{7 r_{12}^7}
\end{equation}
}

$$\underline{\phi^3}$$
\begin{align}
  \text{BIT} &=-\frac{7 (\partial_i\partial_t\phi~\partial_j\partial_t\phi~\partial^i\phi)}{2G\pi}+\frac{\partial^i\partial_t\phi~\partial_j\partial_i\phi~\partial^j\partial_t\phi}{2G\pi}- 3 \frac{\partial_i\partial_j\phi\partial^i\partial^j\phi\partial_t^2\phi}{2 G \pi}  
\end{align}

\hspace{3cm}\begin{tikzpicture}
\draw[postaction={decorate, decoration={markings, mark=at position 1 with {\fill[black] (0,0) circle (2pt);}}}] (1,2) -- (1,1);
    \draw (0,0) -- (2,0);
    \draw (0,2) -- (2,2);
    \draw (1,1) -- (0.5,0);
    \draw (1,1) -- (1.5,0);
     \draw (0,-0.05) -- (2,-0.05);
    \draw (0,2.05) -- (2,2.05);
\end{tikzpicture}\hspace{4cm}
\begin{tikzpicture}{h}
\draw[postaction={decorate, decoration={markings, mark=at position 1 with {\fill[black] (0,0) circle (2pt);}}}] (1,0) -- (1,1);
    \draw (0,0) -- (2,0);
    \draw (0,2) -- (2,2);
    \draw (1.5,2) -- (1,1);
    \draw (0.5,2) -- (1,1);
     \draw (0,-0.05) -- (2,-0.05);
    \draw (0,2.05) -- (2,2.05);
\end{tikzpicture}

{\color{black} A typical vertex factor for the \(\phi^3\) interaction terms comes with a time derivative, which, along with the propagator, is given by:

\begin{equation}
    \mathcal{V} = \frac{(k_1\cdot k_2)^2 \omega_3^2}{k_1^2 k_2^2 k_3^2} \delta^4(k_1+k_2+k_3).
\end{equation}

The $\omega_3$ comes from the time derivative acting on the $e^{i \omega_3 t}$, which itself comes due to $\phi$ being written in the Fourier space, as in Eq. (\ref{eq:fourierphi}), where $k_0 = \frac{\omega}{c}$. The above gravitational interaction leads to the following contributions to the effective Lagrangian:

\begin{align}
I &= \frac{G^2 m_1 m_2^2}{c^4} \bigg[ 
\frac{1}{r_{12}^6} \bigg( 
12 \alpha (v_1^i v_1^i) + 24 \alpha (v_1^i v_2^i) + 12 \alpha (v_2^i v_2^i) 
\bigg) \nonumber\\
&+ \frac{1}{r_{12}^8} \bigg( 
-72 \alpha (v_1^i x^i)^2 - 144 \alpha (v_1^i x^i)(v_2^i x^i) 
- 72 \alpha (v_2^i x^i)^2 
\bigg) \nonumber\\
&+ \frac{1}{r_{12}^6} \bigg( 
12 \alpha (a_1^i x^i) - 12 \alpha (a_2^i x^i) 
\bigg) 
\bigg] \nonumber\\
& + \frac{G^2 m_1^2 m_2}{c^4} \bigg[ 
\frac{1}{r_{12}^6} \bigg( 
12 \alpha (v_1^i v_1^i) + 24 \alpha (v_1^i v_2^i) + 12 \alpha (v_2^i v_2^i) 
\bigg) \nonumber\\
&+ \frac{1}{r_{12}^8} \bigg( 
-72 \alpha (v_1^i x^i)^2 - 144 \alpha (v_1^i x^i)(v_2^i x^i) 
- 72 \alpha (v_2^i x^i)^2 
\bigg) \nonumber\\
&+ \frac{1}{r_{12}^6} \bigg( 
12 \alpha (a_1^i x^i) - 12 \alpha (a_2^i x^i) 
\bigg) 
\bigg].
\end{align}

This result shows how the gravitational interaction modifies the dynamics at different orders, leading to corrections at \( r_{12}^{-5} \) and beyond. The $\omega_3^2$ leads to a velocity or an acceleration term in the results. $x^i$ is the radial vector connecting $m_1$ to $m_2$, \( v_1^i, v_2^i \) are the velocity 3-vectors, and \( a_1^i, a_2^i \) represent acceleration 3-vectors of the two particles.
}
$$\underline{A \phi^2}$$
\begin{align}
    \text{BIT} &= -\frac{3(\partial_i\partial_t\phi~\partial^i\partial^j\phi~\partial_j\partial_kA^k}{4G\pi}-\frac{3(\partial_i\partial_tA^j~\partial^i\partial^k\phi~\partial_j\partial_k\phi)}{4G\pi}+\frac{3(\partial_i\partial^j\phi~\partial^k\partial_kA^{i}\partial_j\partial_t\phi)}{4G\pi} \notag\\
& - \frac{3(\partial^j\partial_iA^{i}\partial_j\partial^k\phi~\partial_k\partial_t\phi)}{4G\pi} +\frac{3(\partial^j\partial_j~A^{i}~\partial_k\partial_t\phi~\partial^k\partial_i\phi)}{4G\pi}-\frac{3(\partial_i\partial_t A_j \partial^k \partial^i\phi~\partial^j\partial_k \phi)}{4G\pi}\notag\\
& +\frac{(\partial_j\partial_t\phi~\partial_k\partial_i\phi~\partial^j\partial^kA^{i})}{4G\pi} +\frac{(\partial_j\partial_i\phi~\partial_k\partial_t\phi~\partial^j\partial^kA^{i}}{4G\pi} -\frac{(\partial^j\partial^i\phi~\partial_i\partial_kA_{j}~\partial^k\partial_t\phi)}{4G\pi}  \notag \\
& - \frac{(\partial^j\partial^i\phi~\partial_j\partial_k A_{i}~\partial^k\partial_t\phi)}{4G\pi}
\end{align}
\hspace{3.5cm}
\begin{center}
    \begin{tikzpicture}
        \draw[dashed, postaction={decorate, decoration={markings, mark=at position 1 with {\fill[black] (0,0) circle (2pt);}}}] (2,1) -- (1,1);
        \draw (0,0) -- (0,2);
        \draw (2,0) -- (2,2);
        \draw (0,1.5) -- (1,1);
        \draw (0,0.5) -- (1,1);
        \draw (-0.05,0) -- (-0.05,2);
        \draw (2.05,0) -- (2.05,2);
    \end{tikzpicture}
    \hspace{1cm}
    \begin{tikzpicture}
        \draw[dashed, postaction={decorate, decoration={markings, mark=at position 1 with {\fill[black] (0,0) circle (2pt);}}}] (0,1) -- (1,1);
        \draw (0,0) -- (0,2);
        \draw (2,0) -- (2,2);
        \draw (-0.05,0) -- (-0.05,2);
        \draw (2.05,0) -- (2.05,2);
        \draw (2,1.5) -- (1,1);
        \draw (2,0.5) -- (1,1);
    \end{tikzpicture}
    \hspace{1cm}
    \begin{tikzpicture}
        \draw[dashed, postaction={decorate, decoration={markings, mark=at position 1 with {\fill[black] (0,0) circle (2pt);}}}] (2,1.5) -- (1,1);
        \draw (0,0) -- (0,2);
        \draw (2,0) -- (2,2);
        \draw (0,1) -- (1,1); 
        \draw (2,0.5) -- (1,1); 
        \draw (-0.05,0) -- (-0.05,2); 
        \draw (2.05,0) -- (2.05,2); 
    \end{tikzpicture}
    \hspace{1cm}
    \begin{tikzpicture}
        \draw[dashed, postaction={decorate, decoration={markings, mark=at position 1 with {\fill[black] (0,0) circle (2pt);}}}] (0,1.5) -- (1,1);
        \draw (0,0) -- (0,2);
        \draw (2,0) -- (2,2);
        \draw (0,0.5) -- (1,1); 
        \draw (2,1) -- (1,1); 
        \draw (-0.05,0) -- (-0.05,2); 
        \draw (2.05,0) -- (2.05,2); 
    \end{tikzpicture}
\end{center}

{\color{black}
The typical vertex factor along with the propagator and momentum conservation is given as follows:
\begin{equation}
    \mathcal{V} = \frac{k_1 \cdot v_j \, k_2^2 \, k_1 \cdot k_3 \, \omega_3}{k_1^2 k_2^2 k_3^2} \delta^4(k_1 + k_2 + k_3) \qquad \qquad j = 1,2
\end{equation}
It contains one time derivative but also includes terms of order \( v_j^i \). As a result, this interaction contributes to either velocity-squared or acceleration dependent terms. The gravitational interaction at this order leads to the following contribution to the effective Lagrangian:  

\begin{align}
    I & = \frac{G^2}{c^4} \bigg[ 
         m_1^2 m_2 \bigg( \frac{ - 24 v_2^2 - 72 (v_1^i v_2^i) + 24 (a_2^i x^i) - 72 (a_1^i x^i)}{r_{12}^6} + \frac{216 (v_1^i x^i) (v_2^i x^i) + 144 (v_2^i x^i )^2+ 216 (v_1^i x^i)^2 }{r_{12}^8} \bigg) \notag \\
        & + m_1 m_2^2 \bigg( \frac{- 24 v_1^2 - 72 (v_1^i v_2^i) + 72 (a_2^i x^i) - 24  (a_1^i x^i)}{r_{12}^6}  \quad + \frac{216 ( v_1^i x^i)( v_2^i x^i) + 216 (v_2^i x^i)^2+ 144 (v_1^i x^i)^2 }{r_{12}^8} \bigg)
    \bigg] \nonumber\\ 
\end{align}

}
$$\underline{\sigma \phi^2}$$
\begin{align}
  \text{BIT} &=-\frac{(\partial^i\partial^j\phi~\partial_i\partial_m\sigma^k_{~k}~\partial^m\partial_j\phi)}{4G\pi}+\frac{3(\partial^i\phi\partial^j\phi~\partial_k\partial_j\phi~\partial^k\partial_i\phi)}{2G\pi}-\frac{(\partial^j\partial^i\phi~\partial_i\partial_m\sigma^k_{~k}~\partial^m\partial_j\phi)}{2G\pi}\notag\\
    &+\frac{3(\partial^j\partial^i\phi~\partial^k\partial_j\phi~\partial_i\partial_m\sigma_{k}^{~m})}{4G\pi}
+\frac{3(\partial^j\partial^i\phi~\partial^k\partial_j\phi~\partial_k\partial_m\sigma_{i}^{~m})}{4G\pi}-\frac{3(\partial^j\partial^i\phi~\partial^k\partial_j\phi~\partial^m\partial_m\sigma_{i k})}{4G\pi}\notag\\
&+\frac{3(\partial^i\partial^m\phi~\partial^k\partial^l\phi~\partial_i\partial_l\sigma_{km})}{2G\pi}-\frac{3(\partial^i\partial^l\phi~\partial^k\partial^m\phi~\partial_i\partial_l\sigma_{km})}{2G\pi}
\end{align}
  \hspace{3.5cm}
\begin{center}
    \begin{tikzpicture}
        \draw[decorate, decoration={snake,amplitude=1,segment length=5pt}, postaction={decorate, decoration={markings, mark=at position 1 with {\fill[black] (0,0) circle (2pt);}}}] (2,1) -- (1,1);
        \draw (0,0) -- (0,2);
        \draw (2,0) -- (2,2);
        \draw (0,1.5) -- (1,1);
        \draw (0,0.5) -- (1,1);
        \draw (-0.05,0) -- (-0.05,2);
        \draw (2.05,0) -- (2.05,2);
    \end{tikzpicture}
    \hspace{1cm}
    \begin{tikzpicture}
        \draw[decorate, decoration={snake,amplitude=1,segment length=5pt}, postaction={decorate, decoration={markings, mark=at position 1 with {\fill[black] (0,0) circle (2pt);}}}] (0,1) -- (1,1);
        \draw (0,0) -- (0,2);
        \draw (2,0) -- (2,2);
        \draw (-0.05,0) -- (-0.05,2);
        \draw (2.05,0) -- (2.05,2);
        \draw (2,1.5) -- (1,1);
        \draw (2,0.5) -- (1,1);
    \end{tikzpicture}
    \hspace{1cm}
    \begin{tikzpicture}
        \draw[decorate, decoration={snake,amplitude=1,segment length=5pt}, postaction={decorate, decoration={markings, mark=at position 1 with {\fill[black] (0,0) circle (2pt);}}}] (2,1.5) -- (1,1);
        \draw (0,0) -- (0,2);
        \draw (2,0) -- (2,2);
        \draw (0,1) -- (1,1); 
        \draw (2,0.5) -- (1,1); 
        \draw (-0.05,0) -- (-0.05,2); 
        \draw (2.05,0) -- (2.05,2); 
    \end{tikzpicture}
    \hspace{1cm}
    \begin{tikzpicture}
        \draw[decorate, decoration={snake,amplitude=1,segment length=5pt}, postaction={decorate, decoration={markings, mark=at position 1 with {\fill[black] (0,0) circle (2pt);}}}] (0,1.5) -- (1,1);
        \draw (0,0) -- (0,2);
        \draw (2,0) -- (2,2);
        \draw (0,0.5) -- (1,1); 
        \draw (2,1) -- (1,1); 
        \draw (-0.05,0) -- (-0.05,2); 
        \draw (2.05,0) -- (2.05,2); 
    \end{tikzpicture}
\end{center}

{\color{black} 

The typical vertex factor for the above diagram contains no time derivatives; however, because of the coupling with the worldline, two factors of velocity come into the vertex factor, which, with the propagator and the momentum conservation, is given by the following
\begin{equation}
    \mathcal{V} = \frac{k_1 \cdot k_2 k_3^2 k_1 \cdot v_j k_2 \cdot v_k}{k_1^2 k_2^2 k_3^2}\delta^4(k_1 +k_2 +k_3) \qquad \qquad j,k = 1,2
\end{equation}
Summing over all possible combinations and Fourier integrating the above leads to the following contribution to the effective Lagrangian

\begin{align}
I &= \frac{G^2 m_1^2m_2}{c^4} \Bigg[ -\frac{12}{r_{12}^6}\Bigg( v_2^2 + 2 v_1^2\Bigg)
-\frac{36}{r_{12}^8}\Bigg((v_2^i x^i)^2+2(v_1^i x^i)^2\Bigg) 
\Bigg] \notag\\
& + \frac{G^2 m_1m_2^2}{c^4}\Bigg[ -\frac{12}{r_{12}^6}\Bigg( v_1^2 +2 v_2^2 \Bigg)
-\frac{36}{r_{12}^8}\Bigg((v_1^i x^i)^2+2(v_2^i x^i)^2\Bigg) 
\Bigg]
\end{align}

}
$$\underline{A^2 \phi}$$
\begin{align}
\text{BIT} &= -\frac{3(\partial_j\partial^k\phi(\partial^m\partial^jA^{i})\partial_i\partial_kA_{m})}{8G\pi} +\frac{3(\partial^i\partial_lA_{m} (\partial^m\partial^lA^{j}) \partial_{i}\partial_{j}\phi)}{8G\pi}-\frac{3((\partial_i \partial_lA^{m})\partial^i\partial^lA^{j} \partial_{m}\partial_{j}\phi)}{8G\pi}\notag\\ &+\frac{3(\partial_i\partial_lA_{m}\partial^l\partial^k A^{i} \partial^{m}\partial_{k}\phi)}{8G\pi}
 \end{align}
 \hspace{3.5cm} \begin{center}
    \begin{tikzpicture}
        \draw[dashed, postaction={decorate, decoration={markings, mark=at position 1 with {\fill[black] (0,0) circle (2pt);}}}] (0,1.5) -- (1,1);
        \draw[dashed, postaction={decorate, decoration={markings, mark=at position 1 with {\fill[black] (0,0) circle (2pt);}}}] (0,0.5) -- (1,1);
         \draw (0,0) -- (0,2);
        \draw (2,0) -- (2,2);
        \draw (1,1) -- (2,1);
        \draw (-0.05,0) -- (-0.05,2);
        \draw (2.05,0) -- (2.05,2);
    \end{tikzpicture}
    \hspace{1cm}
    \begin{tikzpicture}
        \draw[dashed, postaction={decorate, decoration={markings, mark=at position 1 with {\fill[black] (0,0) circle (2pt);}}}] (2,1.5) -- (1,1);
        \draw[dashed, postaction={decorate, decoration={markings, mark=at position 1 with {\fill[black] (0,0) circle (2pt);}}}] (2,0.5) -- (1,1);
         \draw (0,0) -- (0,2);
        \draw (2,0) -- (2,2);
        \draw (0,1) -- (1,1);
        \draw (-0.05,0) -- (-0.05,2);
        \draw (2.05,0) -- (2.05,2);
    \end{tikzpicture}
    \hspace{1cm}
    \begin{tikzpicture}
        \draw[dashed, postaction={decorate, decoration={markings, mark=at position 1 with {\fill[black] (0,0) circle (2pt);}}}] (2,1.5) -- (1,1);
        \draw[dashed, postaction={decorate, decoration={markings, mark=at position 1 with {\fill[black] (0,0) circle (2pt);}}}] (0,1) -- (1,1);
         \draw (0,0) -- (0,2);
        \draw (2,0) -- (2,2);
        \draw (1,1) -- (2,0.5);
        \draw (-0.05,0) -- (-0.05,2);
        \draw (2.05,0) -- (2.05,2);
    \end{tikzpicture}
    \hspace{1cm}
    \begin{tikzpicture}
        \draw[dashed, postaction={decorate, decoration={markings, mark=at position 1 with {\fill[black] (0,0) circle (2pt);}}}] (0,1.5) -- (1,1);
        \draw[dashed, postaction={decorate, decoration={markings, mark=at position 1 with {\fill[black] (0,0) circle (2pt);}}}] (2,1) -- (1,1);
         \draw (0,0) -- (0,2);
        \draw (2,0) -- (2,2);
        \draw (0,0.5) -- (1,1);
        \draw (-0.05,0) -- (-0.05,2);
        \draw (2.05,0) -- (2.05,2);
    \end{tikzpicture}
\end{center}

{\color{black}
A typical vertex factor, propagator, and momentum conservation is given by the following form
\begin{equation}
 \mathcal{V} = \frac{k_1 \cdot k_2 \, k_1 \cdot k_3 \, k_3 \cdot v_j \, k_2 \cdot v_k}{k_1^2 k_2^2 k_3^2}\delta^4(k_1+k_2+k_3)   \qquad \qquad j,k = 1,2
\end{equation}
The above contains two velocity factors, leading to the following contribution to the effective Lagrangian

\begin{equation}
\begin{split}
I &= \frac{G^2}{c^4} \Bigg[ m_1^2 m_2 \Bigg( \frac{96 (v_1^i v_2^i)}{r_{12}^6}  - \frac{144 (v_1^i x^i )(v_2^i x^i)}{r_{12}^8} - \frac{12 v_1^2}{r_{12}^6} - \frac{36 (v_1^i x^i)^2}{r_{12}^8} \Bigg) \nonumber\\ 
& + m_1 m_2^2 \Bigg( \frac{96 (v_1^i v_2^i)}{r_{12}^6} - \frac{144 (v_1^i x^i )(v_2^i x^i)}{r_{12}^8} - \frac{12 v_2^2}{r_{12}^6} - \frac{36 (v_2^i x^i)^2}{r_{12}^8} \Bigg) \Bigg]
\end{split}
\end{equation}

}
    
\hspace{2cm}\underline{Scalar propagator expanded PN 1 $\phi^3$ appearing at PN 2}\\

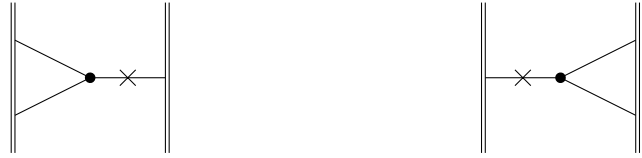
\hspace{3cm} \begin{tikzpicture}
    \draw[postaction={decorate, decoration={markings, mark=at position 1 with {\fill[black] (0,0) circle (2pt);}}}](0,0.5) -- (1,1);
     \draw (0,0) -- (0,2);
    \draw (2,0) -- (2,2);
    \draw (0,1.5) -- (1,1);
    \draw (-0.05,0) -- (-0.05,2);
    \draw (2.05,0) -- (2.05,2);
    \draw[postaction={decorate, decoration={markings, mark=at position 0.5 with {\draw[black] (-3pt,-3pt) -- (3pt,3pt) (-3pt,3pt) -- (3pt,-3pt);}}}] (1,1) -- (2,1);
\end{tikzpicture}\hspace{4cm}
\begin{tikzpicture}
    \draw[postaction={decorate, decoration={markings, mark=at position 1 with {\fill[black] (0,0) circle (2pt);}}}](2,0.5) -- (1,1);
     \draw (0,0) -- (0,2);
    \draw (2,0) -- (2,2);
    \draw (2,1.5) -- (1,1);
    \draw (-0.05,0) -- (-0.05,2);
    \draw (2.05,0) -- (2.05,2);
    \draw[postaction={decorate, decoration={markings, mark=at position 0.5 with {\draw[black] (-3pt,-3pt) -- (3pt,3pt) (-3pt,3pt) -- (3pt,-3pt);}}}] (0,1) -- (1,1);
\end{tikzpicture}\\
{\color{black} Like the $\phi^3$ calculations at 1 PN in Sec. \ref{subsec:1PNR^3}, integrals corresponding to the above diagrams evaluate to zero.}

We obtain the final result by summing over all the previous contributions at 2 PN, leading to a change in the two-body effective potential energy due to the Riemann cubed term in the bulk gravitational action compared to GR. At 2 PN, the change is given as follows:
\begin{align}\label{eq:VeffCubic}
V_{eff,\alpha} = & -\frac{24 G^2 m_1 m_2}{r_{12}^6} \bigg\{ m_1 \bigg[3 (a_1 \cdot r_{12}) - (a_2 \cdot r_{12})\bigg] - m_2 \bigg[ 3 (a_2 \cdot r_{12})- (a_1 \cdot r_{12})\bigg]\bigg\} \notag\\
&- \frac{36 G^2 m_1 m_2 (m_1 + m_2)}{r_{12}^6} \left(v_1^2 + v_2^2 - 2 v_1 \cdot v_2 \right) + \frac{160 \, G^3 \, m_1 \, m_2 \left( m_1^2 + m_2^2 \right)}{7 r_{12}^7} \notag\\
& + \frac{108 \, G^2 \, m_1 \, m_2 \left( m_1 + m_2 \right)}{r_{12}^8} \left( (v_1 \cdot r_{12}) - (v_2 \cdot r_{12}) \right)^2  +\mathcal{O}\left(\frac{1}{r_{12}^{11}}\right)
\end{align}
where $r_{12}$ is the inter orbital separation, $v_i$ for $i = 1, 2$ are the velocities of BH 1 and BH 2 respectively, and $a_i$ for $i = 1, 2$ are the accelerations of BH 1 and 2 respectively. Eq. (\ref{eq:VeffCubic}) can be rewritten in a form where, instead of individual accelerations and velocities, we use relative values. The form is given as follows
\begin{eqnarray}\label{eq:VeffCubicRel}
    V_{eff,\alpha} &=& -\frac{24 G^2 \mu M}{r_{12}^6} \bigg[ 2 \delta M \Vec{a}_{CM} - 4 \mu \Vec{a}_{rel} - M \Vec{a}_{rel} \bigg] \cdot \Vec{r}_{12} - \frac{36 G^2 \mu M^2}{r_{12}^6} v_{rel}^2 \nonumber\\
    && + \frac{108 \, G^2 \, \mu M^2}{r_{12}^8} \left( \Vec{v}_{rel} \cdot \Vec{r}_{12} \right)^2 + \frac{160 \, G^3 \, \mu M^2  \left( M - 2 \mu \right)}{7 r_{12}^7} +\mathcal{O}\left(\frac{1}{r_{12}^{11}}\right)
\end{eqnarray}
where $\frac{1}{\mu} = \frac{1}{m_1}+\frac{1}{m_2}$, $M = m_1 + m_2$, $\delta = \frac{m_1 - m_2}{M}$, $\Vec{a}_{CM} = \frac{m_1 \Vec{a}_1 + m_2 \Vec{a}_2}{M}$, $\Vec{a}_{rel} = \Vec{a}_2 - \Vec{a}_1$, and $\Vec{v}_{rel} = \Vec{v}_2 - \Vec{v}_1$.

{\color{black} Similar to the quadratic Ricci scalar modified effective potential, the above result is also gauge dependent. The corresponding gauge invariant total binding Energy for circular orbits was found to be the following
\begin{eqnarray}
    E_{R^3} &=& -\frac{\mu c^2 x}{2} \left\{ 1 + \left( - \frac{3}{4} -
        \frac{\nu}{12} \right) x + \left( - \frac{27}{8} +
        \frac{19}{8} \nu - \frac{\nu^2}{24} \right) x^2 - \beta \left(\frac{1672}{21} + \frac{7744 \nu}{21}\right) x^6 \right\} \nonumber\\
        \\
    \beta &=& \alpha \left( \frac{G M}{c^2} \right)^{-4} \qquad \qquad x = \left(\frac{G M \Omega}{c^3}\right)^{2/3}, \label{xdef1}
\end{eqnarray}
Derivation of the above has been given in Appendix \ref{app:GIBE}. The above result implies that the actual deviation from GR happens at the sixth PN order, and is consistent with the result of Eq. (2.20) of Ref. \cite{Brandhuber:2019qpg} since at the leading order $x \propto \frac{1}{r_{12}}$. However, our result has a discrepancy with Eq. (A25) of Ref. \cite{Liu:2024atc}. Although both the results are at 6 PN, the coefficient of the 6 PN term differs in our results and theirs. A possible reason could be that Ref. \cite{Liu:2024atc} uses 1PN results, while our study uses results up to 2PN.

}

\section{Conclusions and Discussion}\label{sec:concanddiss}
This study explores the two-particle effective potential energy for two modified gravity theories: quadratic gravity and cubic gravity. We integrated the gravitational degrees of freedom using the Worldline Effective Field Theory (WEFT) approach, employing a Kol-Smolkin metric decomposition to simplify the interactions. The analysis showed that higher-order curvature terms, specifically the particular cubic Riemann tensor contraction, modify the gravitational binding potential at short distances, deviating from the predictions of General Relativity (GR).\\

{\color{black}We observed that the quadratic gravity theory leads to no change in the effective action compared to GR after integrating out the gravitational degrees of freedom, owing to some field redefinitions that keep the structure of the bulk action the same as in GR.} Meanwhile, the cubic gravity term contributed a correction that scaled inversely with the fifth power of the separation. These results suggest that higher-order curvature corrections could significantly alter short-distance gravitational dynamics, potentially providing a means to test these theories in highly compact systems, such as binary black hole mergers. {\color{black} However, the potentials being gauge-dependent cannot be treated as observables. Hence, a gauge-invariant calculation for the total binding energy of a circular orbit was required, which leads to the conclusion that for the cubic modification, the leading-order change in the potential was actually a gauge artefact. For the cubic Riemann case, the actual change happens at the sixth PN order, as was inferred from the gauge invariant binding energy for circular orbits.}\\

The cubic gravity modification, particularly, is noteworthy due to its connection with the renormalization program and field redefinition strategies in quantum gravity. Both approaches independently arrive at the same Riemann cubic term, implying that this term may play a fundamental role in extending GR. Future work should explore this connection further to clarify the implications for quantum gravity theories.

\textcolor{black}{The present analysis underscores how the WEFT framework offers a more streamlined and systematic approach to gravitational dynamics than the traditional PN expansion. Whereas PN calculations have evolved over decades to reach high perturbative orders, the WEFT method achieves comparable accuracy through a compact and transparent field-theoretic formulation that integrates out gravitational degrees of freedom using Feynman diagrams. This efficiency facilitates rapid extensions to higher-curvature corrections and modified gravity scenarios while preserving a clear distinction between conservative and radiative sectors. The demonstrated capability of WEFT to handle complex gravitational interactions with speed and clarity suggests its growing relevance for precision studies of relativistic binaries and theoretical explorations beyond General Relativity.}

\subsection{Future directions}
While this paper focused on the conservative effects of modified gravity theories, radiative effects also play a crucial role, especially in systems that emit gravitational waves. In a follow-up study, we will investigate the radiative corrections to the effective potential and calculate the leading-order changes in gravitational waveforms due to these higher-order curvature terms. This would be an important step in understanding how modified gravity theories might manifest in astrophysical observations, particularly in the context of gravitational wave detections from binary mergers.

\section{Acknowledgments}
We have partially used the package \textbf{EFTofPNG} \cite{Levi:2017kzq} in our calculations. We are indebted to Raj Patil and Jan Steinhoff for important discussions and help regarding the \textbf{EFTofPNG} package. {\color{black}  We also want to thank Guillaume Faye for helping us calculate the relativistic centre of mass as a conserved quantity of a PN Lagrangian till 2 PN.}

\newpage

\appendix

{\color{black}


}
\section{Extra diagrams for cubic gravity at 2 PN}
There were eight diagrams that the \textbf{EFTofPNG} package could not solve. {\color{black} These eight diagrams, even after letting the calculation for the \emph{main} file run for several days, did not compute the results nor gave any errors. We believe that the number of terms in calculating the inverse Fourier transform of the amplitudes of the eight diagrams using the results of the \emph{Nloop} file is computationally expensive for an average machine.} We have therefore manually calculated these terms and only tried to gauge the falloff behaviour of some terms. We restricted ourselves to some terms since there was a dramatic proliferation of terms. We found the following\\

\hspace{7cm}\underline{$\phi^3$ $2^{nd}$ order}
\\

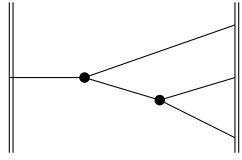
\hspace{6.5cm}\begin{tikzpicture}
\draw[postaction={decorate, decoration={markings, mark=at position 1 with {\fill[black] (0,0) circle (2pt);}}}] (0,1) -- (1,1);
\draw[postaction={decorate, decoration={markings, mark=at position 1 with {\fill[black] (0,0) circle (2pt);}}}] (1,1) -- (2,0.7);
    \draw (0,0) -- (0,2);
    \draw (3,0) -- (3,2);
    \draw (0.05,0) -- (0.05,2);
    \draw (3.05,0) -- (3.05,2);
    \draw (1,1) -- (3,1.7);
    \draw (2,0.7) --  (3,1);
    \draw (2,0.7) --  (3,0.2);
    \end{tikzpicture}\\
    
 \begin{equation}
     \begin{split}
    V_{eff,\alpha} &\propto \int_{k_1...k_5}d^3k_1...d^3k_5 \left[\frac{k_1^2~
    (k_2.k_3)^2~k_3^2(k_4.k_5)^2}{k_1^2~k_2^2~k_3^2~k_4^2~k_5^2}\right]e^{ik_1.x_1}e^{-i(k_2+k_3+k_5).x_2}\delta^3(k_1-k_2-k_3)\delta^3(k_3-k_4-k_5)\\
    &\text{integrate over $k_1$ and $k_3$}\\
    &=\int_{k_2k_4k_5}d^3k_2~d^3k_4~d^3k_5  \left[\frac{((k_2.(k_4+k_5))^2~(k_4.k_5)^2}{k_2^2~k_4^2~k_5^2}\right]e^{i(k_2+k_4+k_5).r}\\
    &=\int_{k_2k_4k_5} d^3k_2~d^3k_4~d^3k_5\left[\left(\frac{(k_2.k_4)^2(k_4.k_5))^2}{k_2^2~k_4^2~k_5^2}\right)+(k4\Leftrightarrow k_5)+2\left(\frac{(k_2.k_4)(k_2.k_5)(k_4.k_5)}{k_2^2~k_4^2~k_5^2}\right)\right] \nonumber\\
    &\times e^{i(k_2+k_4+k_5).r}\\
     \end{split}
 \end{equation}
 $$\propto \frac{G^3m_1m_2^3}{r^{11}}$$
Integrals are done by starting with a simple integral ($I$) and taking its derivative with respect to $r$, where
\begin{eqnarray} 
I &=& \int_k d^3k \frac{e^{ik.r}}{k^2}\propto \frac{1}{r}
\end{eqnarray}
Similarly, we have other diagrams like\\

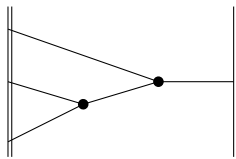
\hspace{2cm}
  \noindent
  \begin{tikzpicture}
\draw[postaction={decorate, decoration={markings, mark=at position 1 with {\fill[black] (0,0) circle (2pt);}}}] (2,1) -- (1,0.7);
\draw[postaction={decorate, decoration={markings, mark=at position 1 with {\fill[black] (0,0) circle (2pt);}}}] (3,1) -- (2,1);
    \draw (0,0) -- (0,2);
    \draw (3,0) -- (3,2);
    \draw (0.05,0) -- (0.05,2);
    \draw (3.05,0) -- (3.05,2);
    \draw (2,1) -- (0,1.7);
    \draw  (1,0.7) --  (0,1);
    \draw  (1,0.7) --  (0,0.2);
    \end{tikzpicture}\hspace{1cm}
$\propto \frac{G^3m_1^3m_2}{r^{11}}$
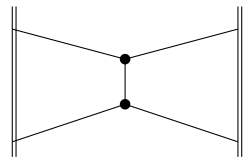
\hspace{1cm}
\begin{tikzpicture}
\draw[postaction={decorate, decoration={markings, mark=at position 1 with {\fill[black] (0,0) circle (2pt);}}}] (0,0.2) -- (1.5,0.7);
\draw[postaction={decorate, decoration={markings, mark=at position 1 with {\fill[black] (0,0) circle (2pt);}}}] (0,1.7) -- (1.5,1.3);
    \draw (0,0) -- (0,2);
    \draw (3,0) -- (3,2);
    \draw (0.05,0) -- (0.05,2);
    \draw (3.05,0) -- (3.05,2);
  \draw (1.5,0.7) -- (3,0.2);
    \draw  (1.5,1.3)--  (3,1.7);
    \draw  (1.5,1.3)--  (1.5,0.7);
    \end{tikzpicture}\hspace{1.5cm}
    $\propto \frac{G^3m_1^2m_2^2}{r^{11}}$\\

\hspace{5cm}\begin{tikzpicture}
\draw[postaction={decorate, decoration={markings, mark=at position 1 with {\fill[black] (0,0) circle (2pt);}}}] (3,1.7) -- (2,1);
\draw[postaction={decorate, decoration={markings, mark=at position 1 with {\fill[black] (0,0) circle (2pt);}}}]  (0,0.3)-- (1,1); 
    \draw (0,0) -- (0,2);
    \draw (3,0) -- (3,2);
    \draw (0.05,0) -- (0.05,2);
    \draw (3.05,0) -- (3.05,2);
    \draw  (1,1) --  (2,1);
    \draw  (2,1)--  (3,0.3);
    \draw  (1,1)--  (0,1.7);
\end{tikzpicture}\hspace{1.5cm}
$\propto \frac{G^3m_1^2m_2^2}{r^{11}}$\\
It was found that there are a total of 4 combinations for 2nd order $\phi^3$ diagrams. Only one case is considered in detail here, but all of them will either vanish or give the same fall of as $r^{-11}$.\\

\hspace{7cm}\underline{$\sigma \phi^2$ $2^{nd}$ order}
\\

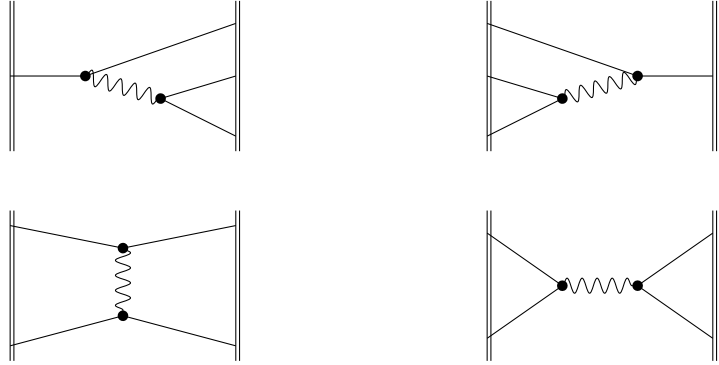
\hspace{3cm}
\begin{tikzpicture}
\draw [decorate, decoration={snake, amplitude=1mm, segment length=2mm}] (1,1) -- (2,0.7);
\draw[postaction={decorate, decoration={markings, mark=at position 1 with {\fill[black] (0,0) circle (2pt);}}}] (0,1) -- (1,1);
\draw[postaction={decorate, decoration={markings, mark=at position 1 with {\fill[black] (0,0) circle (2pt);}}}]  (3,1) -- (2,0.7);
\draw (0,0) -- (0,2);
    \draw (3,0) -- (3,2);
    \draw (0.05,0) -- (0.05,2);
    \draw (3.05,0) -- (3.05,2);
    \draw (1,1) -- (3,1.7);
    \draw (2,0.7) --  (3,0.2);
    \end{tikzpicture}
    \hspace{3cm}
 \begin{tikzpicture}
\draw[postaction={decorate, decoration={markings, mark=at position 1 with {\fill[black] (0,0) circle (2pt);}}}] (0,1)--(1,0.7);
\draw[postaction={decorate, decoration={markings, mark=at position 1 with {\fill[black] (0,0) circle (2pt);}}}] (3,1) -- (2,1);
\draw [decorate, decoration={snake, amplitude=1mm, segment length=2mm}] (2,1) -- (1,0.7);
\draw (0,0) -- (0,2);
    \draw (3,0) -- (3,2);
    \draw (0.05,0) -- (0.05,2);
    \draw (3.05,0) -- (3.05,2);
    \draw (2,1) -- (0,1.7);
    \draw  (1,0.7) --  (0,0.2);
    \end{tikzpicture}\\
    
    \hspace{3cm}
    \begin{tikzpicture}
\draw[postaction={decorate, decoration={markings, mark=at position 1 with {\fill[black] (0,0) circle (2pt);}}}] (0,0.2) -- (1.5,0.6);
\draw[postaction={decorate, decoration={markings, mark=at position 1 with {\fill[black] (0,0) circle (2pt);}}}] (0,1.8) -- (1.5,1.5);
\draw [decorate, decoration={snake, amplitude=1mm, segment length=2mm}] (1.5,1.5) -- (1.5,0.6);
    \draw (0,0) -- (0,2);
    \draw (3,0) -- (3,2);
    \draw (0.05,0) -- (0.05,2);
    \draw (3.05,0) -- (3.05,2);
  \draw (1.5,0.6) -- (3,0.2);
    \draw  (1.5,1.5)--  (3,1.8);
    \end{tikzpicture}
    \hspace{3cm}
    \begin{tikzpicture}
\draw[postaction={decorate, decoration={markings, mark=at position 1 with {\fill[black] (0,0) circle (2pt);}}}] (3,1.7) -- (2,1);
\draw[postaction={decorate, decoration={markings, mark=at position 1 with {\fill[black] (0,0) circle (2pt);}}}]  (0,0.3)-- (1,1); 
\draw [decorate, decoration={snake, amplitude=1mm, segment length=2mm}](1,1) --  (2,1); 
    \draw (0,0) -- (0,2);
    \draw (3,0) -- (3,2);
    \draw (0.05,0) -- (0.05,2);
    \draw (3.05,0) -- (3.05,2);
    \draw  (2,1)--  (3,0.3);
    \draw  (1,1)--  (0,1.7);
\end{tikzpicture}
\begin{equation}
    \nabla^2 \sigma_{ij} \partial_i\partial_k\phi \partial^k\partial^j\phi
\end{equation}
which in momentum space becomes $ k_1^2\sigma(k_1)_{ij}(k_2 \cdot k_3) k_2^ik_3^j$. Thus, we have shown that one of the interactions is very similar to the $2^{nd}$ order $\phi^3$ term. As a result, the calculations for the interactions involving the $2^{nd}$ order $\sigma \phi^2$ closely mirror the $2^{nd}$ order $\phi^3$ interactions, with the main difference being how the $\sigma$ field is incorporated into the diagrams. In $2^{nd}$ order $\sigma \phi^2$, the $\sigma$ field couples quadratically to $\phi$, which alters the diagram structure but does not affect the falloff of the interaction. The power law for distance dependence remains $\frac{1}{r^{11}}$, and the mass dependence in both cases is identical. Thus, while the couplings and diagrams may differ, the long-distance behavior and mass dependence follow the same pattern as in the $2^{nd}$ order $\phi^3$ theory. Hence, we have ignored the $2^{nd}$ order $\phi^3$ and $\sigma \phi^2$ contributions to the binding potential at 2 PN.

{\color{black}

\section{Center of mass for cubic theories}\label{app:CoM}

\subsection{Center of mass as a conserved quantity}

Following Ref. \cite{Bernard:2017ktp}, one notices that the effective Lagrangian obtained from a Lorentz invariant gravitational Lagrangian is also itself Lorentz invariant in a PN fashion. Consider the following infinitesimal Lorentz transformation in a general direction with a boost 3-velocity $\beta^i$
\begin{equation}
        \delta_\beta x^i = \beta^i\,t \qquad \delta_\beta t = \frac{(x^i \beta_i)}{c^2} \label{inf_Lor_trans}
\end{equation}
The above leads to the trajectories $r^i_{j}(t)$ for $j = 1,\,2$ transforming as
\begin{eqnarray}
        \delta_\beta r_j^i &=& \beta^i t - \frac{(\beta^k r_{k,j}) v_j^i}{c^2} \label{traj_trans}\\
        \delta_\beta v_j^i &\equiv& \frac{d}{dt} (\delta_\beta r_j^i) = \beta^i - \frac{(\beta^k v_{k,j}) v^i}{c^2} - \frac{(\beta^k r_{k,j})a^i}{c^2} \label{vel_trans}\\
        \delta_\beta a_j^i &\equiv& \frac{d^2}{dt^2} (\delta_\beta r_j^i) = \mathcal{O}\bigl(\frac{1}{c^2}\bigr) \label{accel_trans}
\end{eqnarray}
For Lagrangians without accelerations, the above transformations corresponds to a changed Lagrangian, which is a total derivative away from the original Lagrangian. The total derivative part or the change in the Lagrangian is given by the following
\begin{equation}
    \delta_\beta L \equiv \frac{d}{dt} \left\{ \beta_i G^i - \sum\limits_{j = 1, 2} \left[ \frac{1}{c^2} (p_j^i v_{i,j}) (\beta_k r_j^k) \right] \right\} \label{deltaLR2}
\end{equation}
where
\begin{equation}
    p_j^i \equiv \frac{\partial L}{\partial v_{i,j}}, \label{momdef}
\end{equation}
and $G^i$ is the \emph{relativistic} center of mass vector.

For Lagrangians with accelerations, Eq. (\ref{deltaLR2}) changes in the following manner
\begin{equation}
    \delta_\beta L \equiv \frac{d}{dt} \left\{ \beta_i G^i + \sum\limits_{j = 1,2} \beta_i q_j^i - \sum\limits_{j = 1, 2} \left[ \frac{1}{c^2} (p_j^i v_{i,j}) (\beta_k r_j^k) \right] \right\} \label{deltaLR3}
\end{equation}
where
\begin{eqnarray}
    p^i_j &\equiv& \frac{\partial L}{\partial v_{i,j}} - \frac{d}{dt} \left( \frac{\partial L}{\partial a_{i,j}} \right) \label{momdefa} \\
    q_j^i &\equiv& \frac{\partial L}{\partial a_{i,j}} \label{amomdef}
\end{eqnarray}

\subsection{Modified center of mass for the cubic Riemann theory}
The effective Lagrangian for the cubic theory is given by the following
\begin{equation}
    L = L_N + \frac{1}{c^2} L_{1PN} + \frac{1}{c^4} L_{2PN}
\end{equation}
where
\begin{eqnarray}
    L_N &=& \frac{1}{2} m_1 v_1^2 + \frac{1}{2} m_2 v_2^2 + \frac{G m_1 m_2}{r_{12}} \label{L_Newt2} \\
    L_{1PN} &=& \frac{1}{8} m_1 v_1^4 + \frac{1}{8} m_2 v_2^4 + \frac{G m_1 m_2}{r_{12}} \left[ \frac{3}{2} \left( v_1^2 + v_2^2 \right) - \frac{7}{2} \left(v_1 \cdot v_2\right) - \frac{1}{2} \left(n_{12} \cdot v_1\right) \left(n_{12} \cdot v_2\right) \right] \nonumber\\
    &&- \frac{G^2 m_1 m_2 \left(m_1 + m_2\right)}{2 r_{12}^2} \label{L_1PN_cubic} \\
    L_{2PN} &=& \frac{1}{16} m_1 v_1^6 + \frac{1}{16} m_2 v_2^6 + G m_1 m_2 \left[ \frac{7}{4} (a_2 \cdot v_1) (n_{12} \cdot v_1) - \frac{7}{4} (a_1 \cdot v_2) (n_{12} \cdot v_2) \right.\nonumber\\
    &&+ \frac{1}{8} (a_2 \cdot n_{12}) (n_{12} \cdot v_1)^2 - \frac{1}{8} (a_1 \cdot n_{12}) (n_{12} \cdot v_2)^2 + \frac{7}{8} (a_1 \cdot n_{12}) v_2^2 - \frac{7}{8} (a_2 \cdot n_{12}) v_1^2 \nonumber\\
    && \frac{1}{r_{12}}\left\{ \frac{7}{8} v_1^2 + \frac{7}{8} v_2^2 + \frac{3}{8} (n_{12} \cdot v_1)^2 (n_{12} \cdot v_2)^2 - \frac{7}{8} (n_{12} \cdot v_1)^2 v_2^2 - \frac{7}{8} (n_{12} \cdot v_2)^2 v_1^2 \right.\nonumber\\
    && \left. \left.- 2 v_1^2 (v_1 \cdot v_2) - 2 v_2^2 (v_1 \cdot v_2) + \frac{3}{2} (n_{12} \cdot v_1) (n_{12} \cdot v_2) (v_1 \cdot v_2) + \frac{1}{4} (v_1 \cdot v_2)^2 + \frac{15}{8} v_1^2 v_2^2\right\}\right] \nonumber\\
    && \frac{G^2 m_1 m_2}{r_{12}^2} \left[ m_1 \left\{ \frac{7}{2} (n_{12} \cdot v_1)^2 + \frac{1}{4} v_1^2 - \frac{7}{2} (n_{12} \cdot v_1) (n_{12} \cdot v_2) + \frac{1}{2} (n_{12} \cdot v_2)^2 - \frac{7}{4} (v_1 \cdot v_2) + \frac{7}{4} v_2^2 \right\} \right.\nonumber\\
    && \left. + m_2 \left\{ \frac{1}{2} (n_{12} \cdot v_1)^2 + \frac{7}{4} v_1^2 - \frac{7}{2} (n_{12} \cdot v_1) (n_{12} \cdot v_2) + \frac{7}{2} (n_{12} \cdot v_2)^2 - \frac{7}{4} (v_1 \cdot v_2) + \frac{1}{4} v_2^2 \right\} \right] \nonumber\\
    && + \frac{G^3 m_1 m_2}{r_{12}^3} \left( \frac{m_1^2}{2} + \frac{19 m_1 m_2}{4} + \frac{m_2^2}{2} \right) + \alpha \left[ \frac{G^2 m_1 m_2}{r_{12}^5} \left\{ m_1 \left(-72 (a_1 \cdot n_{12}) + 24 (a_2 \cdot n_{12}) 
    \right.\right.\right.\nonumber\\
    && \left. + \frac{1}{r_{12}} \left[ 108 (n_{12} \cdot v_1)^2 - 36 v_1^2 - 216 (n_{12} \cdot v_1) (n_{12} \cdot v_2) + 108 (n_{12} \cdot v_2)^2 + 72 (v_1 \cdot v_2 - 36 v_2^2)\right]  \right) \nonumber\\
    && + m_2 \left( -24 (a_1 \cdot n_{12}) + 72 (a_2 \cdot n_{12}) + \frac{1}{r_{12}} \left[ 108 (n_{12} \cdot v_1^2) - 36 v_1^2 - 216 (n_{12} \cdot v_1) (n_{12} \cdot v_2) \right.\right.\nonumber\\
    &&\left. \left.\left.\left. + 108 (n_{12} \cdot v_2)^2 + 72 (v_1 \cdot v_2) - 36 v_2^2 \right]  \right)  \right\} + \frac{160 G^3 m_1 m_2}{7 r_{12}^7} \left(m_1^2 + m_2^2\right) \right]
\end{eqnarray}
Since the modification has both velocity and acceleration, the above will have a non-trivial center of mass modification. Upon applying the transformations (\ref{traj_trans})-(\ref{accel_trans}) and utilizing Eq. (\ref{deltaLR3}), one obtains the following center of mass vector modification (refer to Eq. (B2) in Ref. \cite{Bernard:2017ktp} for the GR relativistic center of mass formula)
\begin{equation}
    G^i_{2PN,R^3} = \frac{48 \alpha G^2 m_1 m_2 (m_1-m_2) n^i_{12}}{r_{12}^5} \label{comR3}
\end{equation}

Therefore, the reduced Lagrangian in the center of mass frame is given by
\begin{equation}
    \mathcal{L} = \mathcal{L}_N + \frac{1}{c^2} \mathcal{L}_{1PN} + \frac{1}{c^4} \mathcal{L}_{2PN}
\end{equation}
where
\begin{eqnarray}
    \mathcal{L}_{N} &=& \frac{v^2}{2} + \frac{G M}{r_{12}} \label{LR3COMN} \\
    \mathcal{L}_{1PN} &=& \frac{v^4}{8} - \frac{3\,\nu\,v^4}{8} +
\frac{G M}{r_{12}}\,\left( \frac{\dot{r}^2\,\nu}{2} + \frac{3\,v^2}{2} +
\frac{\nu\,v^2}{2} \right)-\frac{G^2M^2}{2\,r_{12}^2} \label{LR3COM1PN} \\
    \mathcal{L}_{2PN} &=& G M \left[ -\frac{7 \nu a_v \dot{r}}{4} + a_n \left(-\frac{\dot{r}^2 \nu}{8}  + \frac{7 \nu v^2}{8} \right) + \frac{1}{r_{12}} \left\{ \frac{3 \dot{r}^4 \nu^2}{8} + \left( \frac{\nu}{4} - \frac{5 \nu^2}{4} \right) \dot{r}^2 v^2  \right.\right.\nonumber\\
    && \left.\left.+ \left( \frac{7}{8} - \frac{5 \nu}{4} - \frac{9 \nu^2}{8} \right) v^4  \right\} \right] + \frac{G^2 M^2}{r_{12}^2} \left[ \left(\frac{1}{2} + \frac{41 \nu}{8} + \frac{3 \nu^2}{2}\right) \dot{r}^2 \right.\nonumber\\
    &&\left. + \left( \frac{7}{4} - \frac{27 \nu}{8} + \frac{\nu^2}{2} \right) v^2 \right] + \frac{G^3 M^3}{r_{12}^3} \left( \frac{1}{2} + \frac{15 \nu}{4} \right) + \left( \frac{1}{16} - \frac{7 \nu}{16} + \frac{13
     \nu^2}{16} \right) v^6 \nonumber\\
    && + \alpha \left[ \frac{G^2 M^2}{r_{12}^5} \left\{ \left( - 24 - 96 \nu \right) a_n + \frac{1}{r_{12}} \left( 108 \dot{r}^2 - 36 v^2 \right) \right\} + \frac{G^3 M^3}{r_{12}^7} \left( \frac{160}{7} - \frac{320 \nu}{7} \right) \right] \nonumber\\
\end{eqnarray}
where $a^i = a^i_1 - a^i_2$ is the relative acceleration, $a_v = (a \cdot v)$ and $a_n = (a \cdot n_{12})$ are the tangential and radial components of the relative acceleration respectively.

\section{Gauge invariant binding energy for circular orbits in the cubic Riemann theory}\label{app:GIBE}

For Lagrangians with accelerations, the Legendre transformation to the Energy is given by
\begin{eqnarray}
    \mathcal{E} &=& p_\phi \Omega + q_n a_n - \mathcal{L} \label{ER3def}
\end{eqnarray}
where $q^i$  was defined in Eq. (\ref{amomdef}) and $q_n = (q \cdot n_{12})$. The relative acceleration was found to have the following form in the center of mass frame
\begin{eqnarray}
    a^i &=& A n^i + B v^i
\end{eqnarray}
where $A$ and $B$ are PN functions and are given by
\begin{eqnarray}
    A_N &=& -\frac{G M}{r_{12}^2} \\
    A_{1PN} &=& \frac{G M}{r_{12}^2} \left( \frac{3 \dot{r}^2 \nu}{2} - v^2 - 3 \nu v^2 \right) + \frac{G^2 M^2 (4 + 2 \nu)}{r_{12}^3} \\
    A_{2PN} &=& \frac{G M}{r_{12}^2} \left( -\frac{15 \dot{r}^4 \nu}{8} + \frac{45 \dot{r}^4 \nu^2}{8} + \frac{9 \dot{r^2} \nu v^2}{2} - 6 \dot{r}^2 \nu^2 v^2 - 3 \nu v^4 + 4 \nu^2 v^4 \right) \nonumber\\
    && + \frac{G^2 M^2}{r_{12}^3} \left( 2 \dot{r}^2 + 25 \dot{r}^2 \nu + 2 \dot{r}^2 \nu^2 + \frac{13 \nu v^2}{2} - 2 \nu^2 v^2 \right) - \frac{G^3 M^3}{r_{12}^4} \left( 9 + \frac{87 \nu}{4} \right) \nonumber\\
    && +\alpha \left[ \frac{G^2 M^2}{r_{12}^7} \left( -288 \dot{r}^2 - 4608 \dot{r}^2 \nu + 144 v^2 + 576 \nu v^2 \right) - \frac{G^3 M^3}{r_{12}^8} \left( 256 + 640 \nu \right) \right]\nonumber\\
    \\
    B_{1PN} &=& \frac{G M}{r_{12}^2} \left( 4 \dot{r} - 2 \dot{r} \nu \right) \\
    B_{2PN} &=& \frac{G M}{r_{12}^2} \left( -\frac{9 \dot{r}^3 \nu}{2} - 3 \dot{r}^3 \nu^2 + \frac{15 \dot{r} \nu v^2}{2} + 2 \dot{r} \nu^2 v^2 \right) - \frac{G^2 M^2}{r_{12}^3} \left( 2 \dot{r} + \frac{41 \dot{r} \nu}{2} - 4 \dot{r} \nu^2 \right) \nonumber\\
    && \frac{G^2 M^2 \alpha}{r_{12}^7} \left( -144 \dot{r} + 1152 \dot{r} \nu \right)
\end{eqnarray}
We obtain the modified Kepler's third law for the cubic Riemann case by considering $a^i = - \Omega^2 r_{12} n^i_{12}$. We obtain the radial separation as a function of $x$ as
\begin{equation}
    r_{12} = \frac{G M}{c^2} \left[ \frac{1}{x} - 1 + \frac{\nu}{3} + \left( \frac{19 \nu}{4} + \frac{\nu^2}{9} \right) x + \beta \left( \frac{112}{3} + \frac{64 \nu}{3} \right) x^5 \right] \label{rinxR3}
\end{equation}
where we have scaled $\alpha$ as
\begin{equation}
    \alpha = \beta \left(\frac{G M}{c^2}\right)^4
\end{equation}
Therefore, we after substituting Eq. (\ref{rinxR3}) in Eq. (\ref{ER3def}), one obtains the following gauge invariant binding Energy for circular orbits for the cubic Riemann modification
\begin{equation}
    \mathcal{E} = -\frac{c^2 x}{2} \left\{ 1 - \left( \frac{3}{4} +
        \frac{\nu}{12} \right) x + \left( - \frac{27}{8} +
        \frac{19}{8} \nu - \frac{\nu^2}{24} \right) x^2 - \beta \left(\frac{1672}{21} + \frac{7744 \nu}{21}\right) x^6 \right\}
\end{equation}
which clearly shows that the potential at 2 PN was a gauge artefact and the actual modification comes at 6 PN.

}
\input{apssamp.bbl}

\end{document}

%% file: apssamp.bbl
%